\documentclass[letterpaper]{article} 
\usepackage{aaai25}  
\usepackage{times}  
\usepackage{helvet}  
\usepackage{courier}  
\usepackage[hyphens]{url}  
\usepackage{graphicx} 
\usepackage{natbib}  
\usepackage{caption} 
\usepackage{algorithm}
\usepackage{algorithmic}

\usepackage{newfloat}
\usepackage{listings}
\DeclareCaptionStyle{ruled}{labelfont=normalfont,labelsep=colon,strut=off} 
\floatstyle{ruled}
\newfloat{listing}{tb}{lst}{}
\floatname{listing}{Listing}
\usepackage{xcolor}
\newcommand{\answerYes}[1]{\textcolor{blue}{#1}} 
 
\newcommand{\answerNA}[1]{\textcolor{gray}{#1}}

\newcommand{\rr}[1]{{#1}}  
\usepackage{amsmath}
\usepackage{dsfont}
\usepackage{tikz}
\newcommand\github{\url{https://github.com/snap-stanford/llm-social-network}}

\nocopyright

\title{LLMs Generate Structurally Realistic Social Networks\\But Overestimate Political Homophily}
\author{
 Serina Chang\textsuperscript{\rm 1}\equalcontrib,
 Alicja Chaszczewicz\textsuperscript{\rm 1}\equalcontrib,
 Emma Wang\textsuperscript{\rm 1},\\
 Maya Josifovska\textsuperscript{\rm 1,2},
 Emma Pierson\textsuperscript{\rm 3},
 Jure Leskovec\textsuperscript{\rm 1}
}
\affiliations{
 \textsuperscript{\rm 1}Department of Computer Science, Stanford University\\
 \textsuperscript{\rm 2}Department of Computer Science, University of California, Los Angeles\\
 \textsuperscript{\rm 3}Department of Computer Science, Cornell University
}

\begin{document}

\maketitle

\begin{abstract}
Generating social networks is essential for many applications, such as epidemic modeling and social simulations.
\rr{The emergence of generative AI, especially large language models (LLMs), offers new possibilities for social network generation: LLMs can generate networks without additional training or need to define network parameters, and users can flexibly define individuals in the network using natural language. 
However, this potential raises two critical questions: 1) are the social networks generated by LLMs realistic, and 2) what are risks of bias, given the importance of demographics in forming social ties?}
To answer these questions, we develop three prompting methods for network generation and compare the generated networks to a suite of real social networks.
We find that more realistic networks are generated with ``local'' methods, where the LLM constructs relations for one persona at a time, compared to  ``global'' methods that construct the entire network at once.
We also find that the generated networks match real networks on many characteristics, including density, clustering, connectivity, and degree distribution.
However, we find that LLMs emphasize political homophily over all other types of homophily and significantly \textit{overestimate} political homophily compared to real social networks.
\end{abstract}

%
\begin{links}
    \link{Code}{https://github.com/snap-stanford/llm-social-network}
\end{links}

\section{Introduction}
The ability to generate realistic social networks is crucial for many applications, when the true social network cannot be observed (e.g., for privacy reasons) or a realistic network is desired between hypothetical individuals. 
For example, in epidemic modeling, synthetic social networks are frequently used so that researchers can model the spread of disease based on who has come into contact with whom \cite{barrett2009contact,block2020covid}. 
Synthetic networks are also useful for simulating and analyzing social media platforms \cite{perez2015endorsements,sagduyu2018media} and social phenomena, such as polarization and opinion dynamics \cite{dandekar2013polarization,das2014opinion}.

Deep learning approaches to network generation typically require training on many domain-specific networks \cite{you2018graphrnn}, making it difficult to generalize to new settings where networks are not yet observed. 
\rr{On the other hand, some classical network models require far less or no training, since they only have a few parameters, but in exchange for this simplicity, they make rigid and unrealistic assumptions about how social networks form.}
For example, Erdős–Rényi models \rr{only take in two inputs, $n$ (the number of nodes) and $p$, and they assume each edge forms with uniform probability $p$} \cite{erdds1959random}.

In contrast, \rr{LLMs balance these challenges: they can generate social networks in a zero-shot fashion, without any additional training or need to define network parameters, but they can also take in rich, flexible inputs describing individuals in natural language and use those inputs to generate a network between those individuals.
A key question, however, is whether the social networks generated by LLMs are \textit{realistic}.}
On one hand, LLMs have demonstrated capabilities to realistically simulate human responses and interactions \cite{aher2023replicate,park2023generative,argyle2023samples}, suggesting that they may be able to generate realistic social networks as well.
On the other hand, LLMs sometimes struggle with reasoning over graphs \cite{wang23nlgraph,fatemi2024talk} and it is unclear if their language abilities generalize to structured objects like networks, so that they can reproduce structural characteristics of social networks such as low density and long-tailed degree distributions.

Furthermore, a central concern with using LLMs in social settings is bias.
Prior works have shown that LLMs produce stereotyped descriptions of individuals based on their demographics \cite{cheng2023marked,cheng2023compost} and skew towards the liberal opinions \cite{santurkar2023opinions}.
These demographics, such as gender and political affiliation, play essential roles in the formation of real-world social networks, resulting in well-documented demographic homophily \cite{mcpherson2001birds,kossinets2009homophily,halberstam2016homophily}.
Thus, we cannot evaluate whether LLMs' social network generation is realistic without incorporating demographics into our experiments; at the same time, we need to analyze how LLMs reason about these demographic features and investigate potential signs of bias. 

\begin{figure*}[t]
    \centering
    \includegraphics[width=\linewidth]{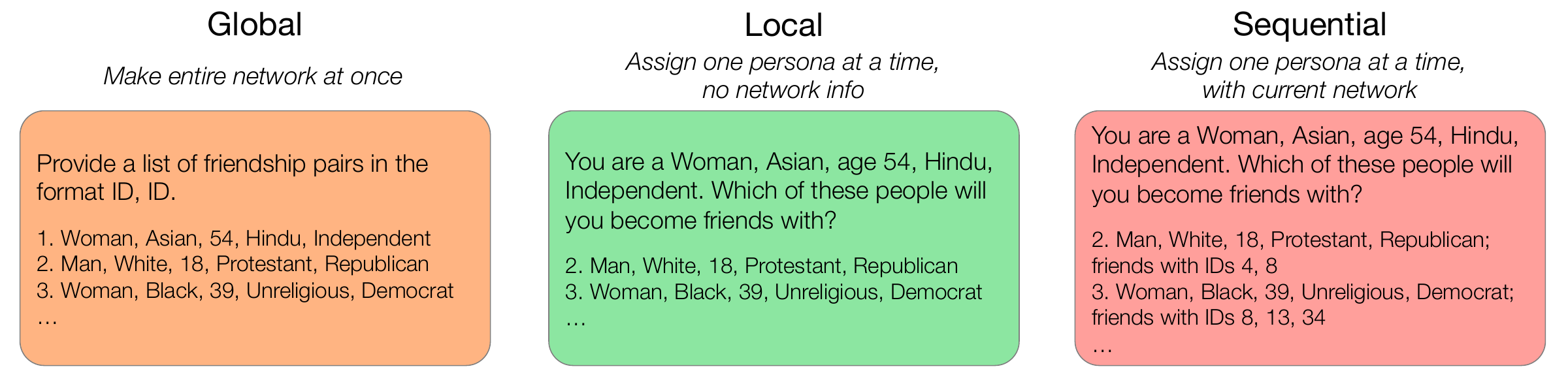}
    \caption{Our three prompting methods to generate social networks with LLMs. See full prompts in Figures~\ref{fig:global-prompt}-\ref{fig:sequential-prompt}.}
    \label{fig:prompts}
\end{figure*}

The goal of this work is to study these two issues---realism and bias---in the context of social network generation with LLMs. Our research questions are as follows:\vspace{8pt}\\
\textbf{RQ1:} Can LLM-generated networks match real-world social networks  on structural characteristics? How do different prompting methods result in different network structures?\vspace{5pt}\\
\textbf{RQ2:} Can LLMs capture demographic homophily? How do levels of homophily vary across demographic variables, and are there signs of bias?\vspace{5pt}\\
\textbf{RQ3:} How does incorporating interests, beyond demographics, affect LLM-generated networks?\vspace{8pt}\\
To answer these questions, we propose three \rr{zero-shot} prompting methods for social network generation (Figure \ref{fig:prompts}).
First, we find that ``local'' methods, where the LLM takes on the perspective of one person at a time, yield more realistic networks than ``global'' methods, where the LLM constructs the entire network at once, even though the LLM receives less information in the local setting.
The LLM is able to generate networks that match real networks on many structural characteristics, including density, clustering, connectivity, and degree distribution.
The LLM also exhibits clear demographic homophily, across gender, age, race/ethnicity, religion, and political affiliation.
However, the LLM consistently emphasizes homophily in political affiliation above all other demographic variables, and overestimates levels of political homophily compared to reported levels in real online \textit{and} offline social networks.
Finally, we find that incorporating LLM-generated interests does not reduce political homophily, since the interests themselves encode political stereotypes. 
Overall, our work demonstrates the promise of using LLMs for social network generation while calling attention to challenges around integrating demographics.
\section{Related Work}
\label{sec:related}
\paragraph{Social simulation with LLMs.}
Prior work has demonstrated LLMs' abilities to realistically simulate human responses and interactions \cite{aher2023replicate,park2023generative,argyle2023samples} \rr{and studied the dynamics of LLM agents interacting in a population, such as how conventions or consensus arise \cite{ashery2024convention,marzo2024consensus}.
However, while simulating interactions over social networks, existing work focuses less on using LLMs to generate the networks themselves, either making simplistic assumptions about the network structure such as sampling agents randomly to interact \cite{park2022social,chuang2023simulating,ashery2024convention} or assuming fully connected networks \cite{marzo2024consensus}}, or requiring human involvement in building the network \cite{gao2023s,zhou2024sotopia}.
To improve the realism and usability of these simulations, it is essential to also explore LLMs' abilities to \textit{generate} the network structure, a prerequisite to simulating interactions over networks.

A few contemporaneous works have explored LLMs for social network generation, with different focuses from ours.
\citet{marzo2023scalefree} focus on degree distribution, showing that scale-free networks emerge from interactions between LLMs.
\citet{he2023chatbot} focus on content homophily, analyzing a simulated society powered by LLM chatbots.
The most similar work to ours is \citet{papachristou2024network}, who analyze whether LLMs demonstrate network formation principles, such as preferential attachment and homophily. 
While their work establishes the existence of \textit{general} network principles, our work compares generated networks to real social networks directly, computing many network metrics, and shows that all metrics can be matched at once, while their experiments primarily explore one principle at a time, with a different prompt for each principle.
Also, to test homophily, they consider hobby, favorite color, and location, while we explore key demographic features. 

\paragraph{LLM social biases.} 
Using LLMs in social contexts raises concerns of biases and stereotyping \cite{cheng2023marked,cheng2023compost,wang2024cannot}.
When responding to public opinion or political questions, LLMs' answers often skew liberal \cite{santurkar2023opinions,hartmann2023political}.
When assigned a persona, LLMs show worse reasoning capabilities when assigned certain demographics \cite{gupta2024reasoning} and produce more toxic content under certain personas \cite{deshpande-etal-2023-toxicity}.
However, bias in the context of social network generation remains unexplored.
In this work, we investigate such biases by studying the effects of demographic variables on LLM-generated social networks.

\paragraph{Graph generation.}
\rr{Deep learning approaches aim to learn graph generation directly from observed data, resulting in realistic generated networks \cite{you2018graphrnn,simonovsky2018graphvae,mlgraphgeneration}. 
However, these models require a set of training graphs, and typically have fixed feature sets, so they cannot be applied in situations without access to real-world network data, and cannot be easily extended to incorporate new features, while LLMs can be.
In contrast, some simple classical models can generate networks with only 1-2 parameters, which means they can generate networks without requiring training and with minimal need to define parameters.
However, their simplicity often results in unrealistic network structures.}
For example, Erdős–Rényi models assume that each edge forms with uniform probability \cite{erdds1959random},
Watts–Strogatz models \cite{watts1998collective} generate small-world networks but struggle to produce realistic degree distributions, and
Barabási–Albert models \cite{barabasi1999emergence} capture power-law degree distributions but miss community structure and clustering.

\paragraph{LLMs \& graphs.}
The use of LLMs has been explored for various graph tasks \cite{li2024surveygraphmeetslarge, jin2024largelanguagemodelsgraphs}, such as graph reasoning \cite{wang23nlgraph,fatemi2024talk}, 
node classification \cite{zhao2023graphtextgraphreasoningtext, chen2024exploring, ye-etal-2024-language}, or tasks on knowledge graphs \cite{Pan_2024}.
These works have shown that LLMs possess preliminary graph reasoning capabilities, but struggle on larger graphs or harder tasks, such as finding a Hamilton path \cite{wang23nlgraph}.
\citet{yao2024exploringpotentiallargelanguage} is one of the first to explore LLMs for graph generation, but they explore it in the context of molecule generation, not social networks.

\section{Generating Social Networks}
\label{sec:prompting-methods}
Our process for generating social networks involves two steps: first, constructing a set of personas, and second, having the LLM generate networks over those personas.
Here, we summarize the two steps, with details in Appendix~\ref{sec:app-methods}.

\subsection{Persona construction}
For each persona, we include their gender, age, race/ethnicity, religion, and political affiliation, which are salient dimensions of homophily in real social networks \cite{mcpherson2001birds,halberstam2016homophily}.
We sample these characteristics based on the distribution of the US population.
Using US Census data \cite{census2023joint}, we acquire the joint distribution for gender, age, and race/ethnicity.
Then, we sample the persona's religion, conditioned on their race/ethnicity \cite{statista2016racereligion,prri2020religious}, and political affiliation, conditioned on their gender and race/ethnicity \cite{pew2024political,brookings2022native}.
In Section \ref{sec:interests}, we also experiment with including interests for each persona, instead of only demographic variables.

\subsection{Network generation}
We design three prompting methods for generating social networks, which we summarize in Figure \ref{fig:prompts}.

\paragraph{Global.}
In our first method, which we call ``Global'', we provide the LLM with the entire list of personas, and prompt it to construct the network between them, in the form of edge pairs (referring to each persona's ID).

\paragraph{Local.}
In our second method, which we call ``Local'', we have the LLM take on the identity of one persona at a time, e.g., by saying, ``You are a Woman, Asian, age 54, Hindu, Independent.''
We provide the LLM with the list of all other personas (in the same format) and prompt it to pick friends for the persona it is currently assigned.
To construct the entire network, we iterate through all personas in a random order, and we keep an edge between personas A and B if the LLM selects B when acting as A or vice versa (so we do not require both to select each other).
This method is inspired by techniques in machine learning that similarly model the graph generation process by iterating through nodes and selecting edges for each node at a time \cite{you2018graphrnn}.

\paragraph{Sequential.}
In our third method, we also assign the LLM one persona at a time, but in addition to providing the list of all other personas, we also provide information about the constructed graph so far.
We experiment with providing each persona's full list of current friends versus only their degree (i.e., total number of friends).
These variations are similar to the preferential attachment experiments in \citet{papachristou2024network}, where they also experimented with providing neighborhood information versus only degree, although their experiments only considered the network and no demographic features.
In contrast, our experiments---and in particular, the comparison of the Local to Sequential methods---reveal how providing demographic and network information compare to only providing demographic information.

\section{Comparison to Real Networks}
\label{sec:metrics}
To evaluate the realism of our generated networks, we gather a set of real social networks from the \rr{CASOS \cite{casos} and KONECT \cite{konect}} repositories.
\rr{We kept networks that described \textit{friendships} between individuals, 
which filtered out other types of networks, such as work-related interactions or visiting ties between families.}
We included eight real networks, which capture \rr{friendships} within diverse communities, such as among physicians, students, and prisoners (see Appendix~\ref{sec:app-structure} for details).

We extract graph-level and node-level metrics from the real networks and our generated networks, and compare their distributions.
For consistency, we treat all networks as undirected.
Since the number of nodes varies across networks, we focus on network metrics that are comparable across graphs of different sizes, and scale those that are dependent on network size based on how they are expected to scale in an Erdös-Rényi random graph \cite{erdds1959random}.
Below, we define and motivate the network metrics that we evaluate on.

\paragraph{Density.} 
A basic property of a network is its density of edges, and social networks tend to be sparse, meaning lower density \cite{wong2006spatial}.
Density computed as the number of observed edges divided by the total number of possible edges in the network, which comes out to $\frac{2E}{N(N-1)}$, where $N$ is the number of nodes and $E$ is the number of edges in the network.

\paragraph{Average clustering coefficient.}
Social networks are known to exhibit clustering, where one's friends are likely to be friends with each other \cite{alizadeh2017spatial}.
For a node $i$, its clustering coefficient is $\frac{2E_i}{N_i(N_i-1)}$, where $N_i$ is its number of neighbors and $E_i$ is the number of edges between its neighbors. 
The average clustering coefficient computes the average over nodes.

\paragraph{Largest connected component (LCC).}
Social networks are known to be well-connected \cite{ugander2011facebook}, with the vast majority (over 99\%) of the nodes in the largest connected component (LCC), i.e., the largest subgraph where all nodes within the subgraph are reachable by each other.
Thus, as a metric, we compute the proportion of all nodes in the network that are in the LCC, $\frac{N_{LCC}}{N}$.

\paragraph{Average shortest path.}
Social networks are not only well-connected, meaning nodes can reach each other, but also they can reach each other in relatively short paths \cite{alizadeh2017spatial}.
So, we measure the average shortest path over all pairs of nodes in the LCC,
divided by $\log N_{LCC}$, since shortest paths scale with $\log N$ in Erdös-Rényi graphs \cite{watts1998collective}.
We compute shortest paths within the LCC instead of the entire network, since the distance between two disconnected nodes is infinite.

\paragraph{Community structure.} 
Social networks exhibit strong community structure, with more edges within communities and fewer edges across communities \cite{newman2004community}.
To measure community structure, first we use the Louvain algorithm \cite{blondel2008fast} to partition the network into communities,
then we assess the quality of the partition with modularity (Eq.~\ref{eqn:modularity}).
Higher levels of community structure correspond to higher modularity.

\paragraph{Degree distribution.} 
Social networks are said to follow a power law degree distribution, where $P(k) \sim k^{-\gamma}$, for degree $k$ and constant $\gamma$ \cite{barabasi1999emergence}.
This results in long-tailed degree distributions with a few people having far more friends than most others.
To measure degree distribution, we compute the degree of each node in the network, and to make degree comparable across graphs, we divide degree by $N$. 
To summarize degree distribution from a set of networks, we pool all of the (normalized) degrees of nodes in the networks in the set, and compute the distribution over the pooled degrees in bins of 0.05, from 0 to 1. 

\paragraph{Homophily.}
Finally, social networks are known to exhibit homophily, where ``birds of a feather flock together'', i.e., people with similar traits are likelier to be friends \cite{mcpherson2001birds}.
\rr{To measure homophily, we use a common metric adopted in prior work \cite{mcpherson2001birds,easley2010networks,smith2014gss}: the ratio of observed-to-expected cross-group edges.
Specifically, first we compute the \textit{observed} proportion of edges that are cross-group (e.g., different gender), then we compute the \textit{expected} proportion of edges that are cross-group (based on the number of nodes that belong to each group), then we take the ratio of these two proportions.
We define this as
\begin{align}
    H &= \frac{C_{\textrm{obs}}}{C_{\textrm{exp}}} = \frac{\frac{\sum_{ij} A_{ij} \cdot \mathds{1}[g_i \neq g_j]}{E}}{\frac{\sum_g \sum_{g' \neq g} N_g N_{g'}}{N (N-1)}}, \label{eqn:cross-group-ratio} 
\end{align}
where $A_{ij}$, as the adjacency matrix, is 1 if nodes $i$ and $j$ are connected and 0 otherwise; $g_i$ indicates node $i$'s group; and $N_g$ is the number of nodes in group $g$.\footnote{For age, we use the average age difference in observed edges divided by the expected age difference (i.e., average difference over all possible pairs of nodes).}
If the ratio is below 1, this indicates homophily, since there are fewer cross-group edges than expected, while ratios above 1 indicate heterophily (e.g., this appears in heterosexual dating networks).}
\section{Results}
\label{sec:results}
\begin{figure}
    \centering
    \includegraphics[width=\linewidth]{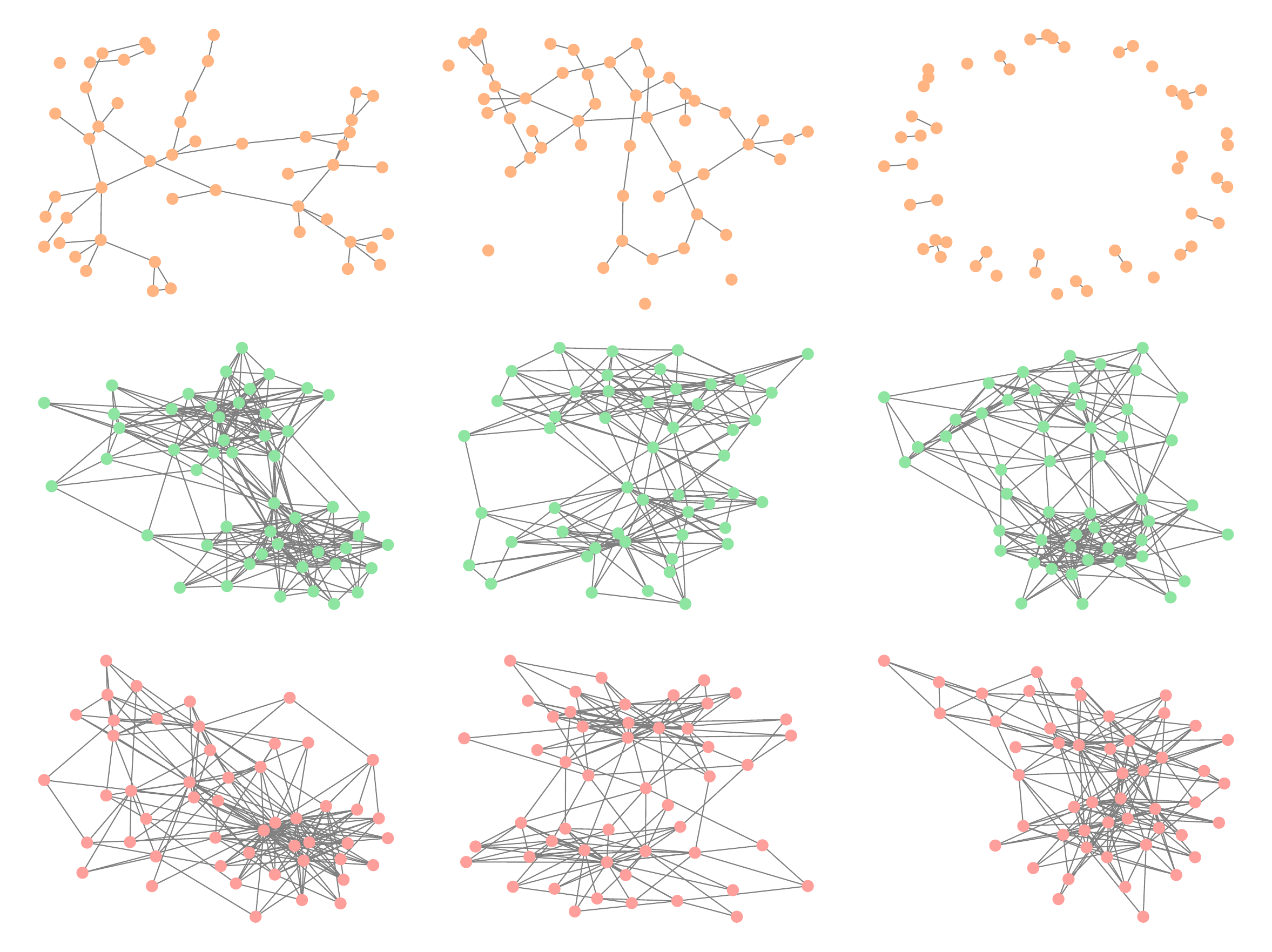}
    \caption{Examples of social networks generated by our three prompting methods: Global (top), Local (middle), and Sequential (bottom).}
    \label{fig:networks}
\end{figure}

\begin{figure*}
    \centering
    \includegraphics[width=\linewidth]{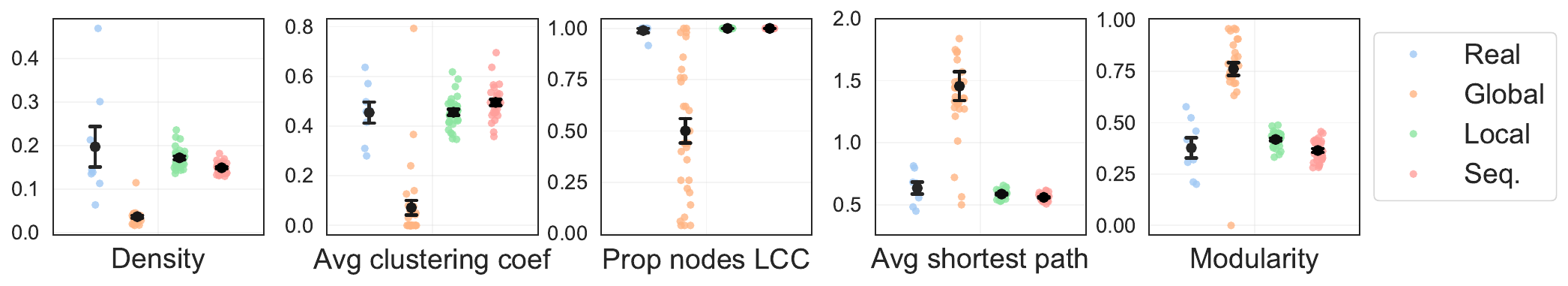}
    \caption{Graph-level metrics over real and generated social networks. We visualize mean and standard error (in black) and individual data points corresponding to each network.}
    \label{fig:metrics}
\end{figure*}

\begin{figure}
    \centering
    \includegraphics[width=\linewidth]{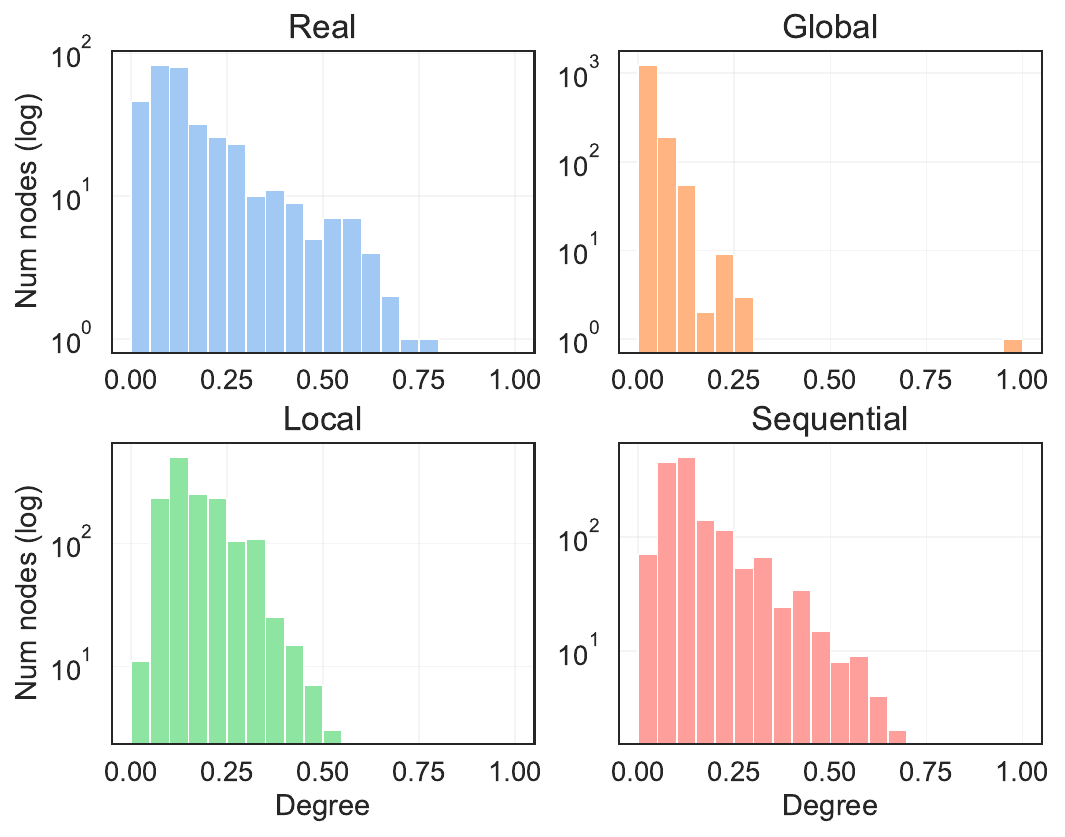}
    \caption{Degree distributions over real and generated social networks. For each set of networks, we pool degrees over nodes in the networks (Section~\ref{sec:metrics}).}
    \label{fig:degree}
\end{figure}

\begin{figure*}
    \centering
    \includegraphics[width=\linewidth]{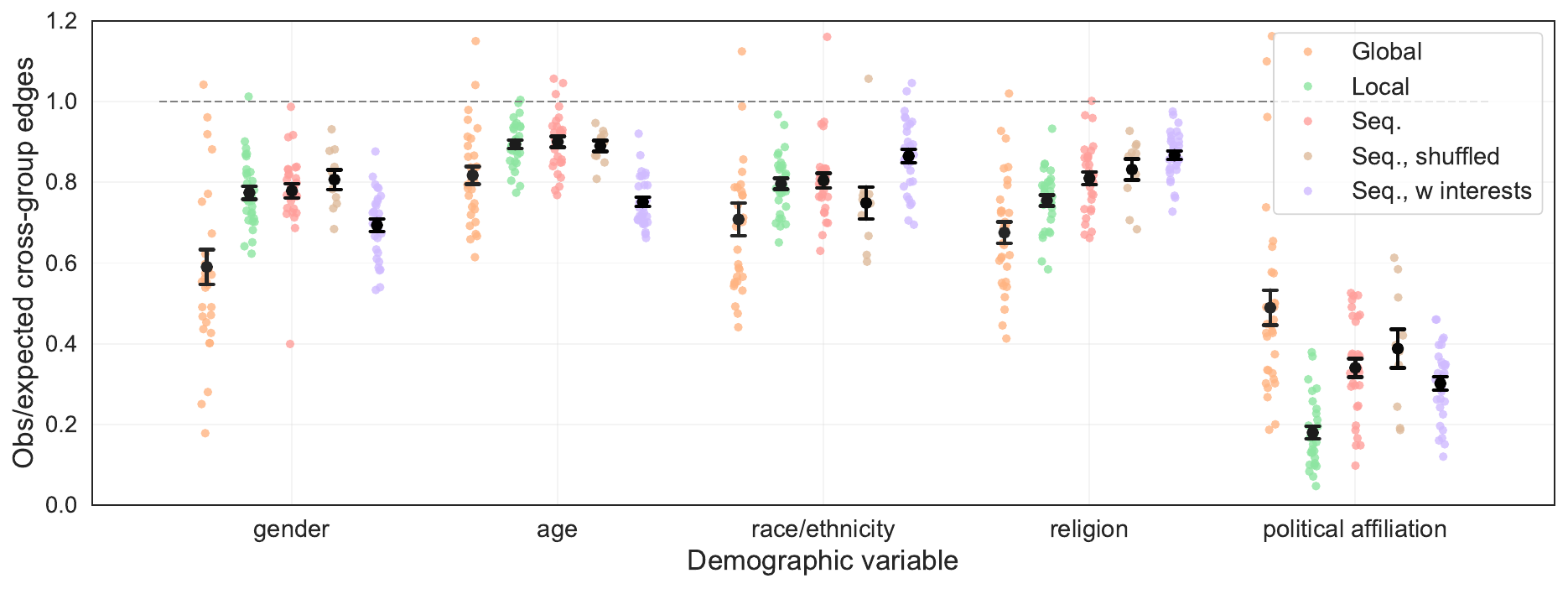}
    \caption{Rates of homophily in our generated networks, per demographic variable. 
    Ratios \rr{below} 1 (marked by the grey line) indicate homophily, with \rr{lower} ratios indicating more homophily.
    \rr{We visualize mean and standard error (in black) and individual data points corresponding to each network.}
    }
    \label{fig:homophily}
\end{figure*}

\paragraph{Experimental set-up.}
We experiment with the following LLMs: OpenAI's GPT-3.5 Turbo and GPT-4o \cite{brown2020gpt3,openai2023gpt4}, Meta's Llama 3.1 (8B and 70B) \cite{llama2023}, and Google's Gemma 2 (9B and 27B) \cite{gemma2024}.
\rr{These six models represent a range across companies, different model sizes, and proprietary (the GPT models) versus open-source (the Llama and Gemma models).}
We find that GPT-3.5 Turbo performs the best at matching the real social networks, so we report results from GPT-3.5 Turbo in the main text, but we report results from all models in Appendix~\ref{sec:diff-llms}.
We show that our main results about political homophily being most emphasized and overestimated hold for all six models.
We also include sensitivity analyses, with different temperatures (Figure~\ref{fig:comparing-temp}) and minor changes to the prompt (Figure~\ref{fig:prompt-change}), and show that results are stable.

We sample $N=50$ personas and we use the same set of personas for all LLM experiments.
In Table~\ref{tab:personas}, we report the demographic make-up of these 50 sampled personas. 
For each prompting method, we generate 30 networks, to capture variation in prompting (e.g., order of personas) and model response.
For the graph-level metrics (all metrics in Section~\ref{sec:metrics} besides degree distribution), 
we visualize their mean and standard error, along with individual data points (Figures~\ref{fig:metrics} and \ref{fig:homophily}).
Visualizing both the standard error and individual data points capture inferential uncertainty and outcome variability, as recommended by \citet{zhang2023illusion}.
For degree (Figure~\ref{fig:degree}), we visualize the pooled degree distribution over nodes in the networks, as described in Section~\ref{sec:metrics}.

\subsection{Evaluating network structure}
\label{sec:structure}
Here, we describe our main results from evaluating the structure of the generated networks.

\paragraph{Local and Sequential are more realistic than Global.}
First, we find that the prompting methods produce visually different network structures, as shown in Figure~\ref{fig:networks}.
Furthermore, the networks produced by the Global method are far less realistic than those produced by Local and Sequential.
As shown in Figures \ref{fig:metrics}-\ref{fig:degree}, Global has unrealistically low density, clustering, and connectivity, too much community structure, and misses the long tail of the degree distribution. 
In comparison, Local and Sequential overlap with the real distributions for all graph-level metrics and show much greater variation in node degrees.

Thus, LLMs produce more realistic social networks when we assign the LLM to act as one persona at a time, instead of prompting it to produce the entire network at once.
This is interesting, since the LLM has strictly less information under the one-persona setting: in the Local setting, it has no access to any network information, only making local decisions per persona based on demographics, and in the Sequential setting, it only knows the network based on previous personas' choices without any ability to see into the future.
In comparison, the Global method allows the LLM to take into account the entire network at once, along with all personas' demographic information, so that it can theoretically consider dependencies between all these pieces of information.
However, the LLM is not able to effectively leverage all of these dependencies and instead produces far less realistic networks.

\paragraph{Sequential captures long-tail degree distribution.}
By comparing Local and Sequential, we can isolate the impact of incorporating network information, on top of providing demographic information.
The main difference between the two methods is the degree distribution: as shown in Figure~\ref{fig:degree}, Sequential gets much closer to the real degree distribution, in terms of exhibiting a long tail, while Local approximately matches the mode without matching the tail.
We also find that Sequential exhibits slightly less community structure and less homophily than Local, which makes sense since Local only matches on demographic similarity while Sequential also takes into account network information.

Thus, the Sequential method is able to match real social networks on many structural characteristics.
In Table~\ref{tab:metrics}, we quantify how well each LLM method matches the real networks, and compare to classical network models \rr{with few parameters}, including random graph models \cite{erdds1959random}, small-world models \cite{watts1998collective}, and preferential attachment models \cite{barabasi1999emergence}.
Even in the best case---when we allow the classical models to choose parameters based on the real networks---these models cannot match all of the characteristics as well as our Local or Sequential methods can. 
For example, when using the Kolmogorov-Smirnov statistic (Eq.~\ref{eqn:ks}) to measure the distance between the generated versus real networks' distributions per structural characteristic, Sequential achieves an average distance of 0.330, while the best classical model (small-world) yields an average distance of 0.499, 51\% higher than Sequential.
\rr{Thus, the LLM achieves better realism than these classical network models when generating social networks, while maintaining their advantages of requiring no additional training or minimal need to define parameters.}

\subsection{Evaluating homophily}
We measure homophily for gender, \rr{age}, race/ethnicity, religion, and political affiliation, using \rr{the ratio of observed-to-expected proportion of cross-group edges} (Eq. \ref{eqn:cross-group-ratio}).

\paragraph{LLMs capture homophily, with greatest emphasis on politics.}
In Figure~\ref{fig:homophily}, we show that, across all prompting methods and demographic variables, \rr{the ratio is significantly below 1, indicating that the generated networks clearly exhibit homophily}.
Furthermore, we see different levels of homophily for different demographic variables.
For the more realistic Local and Sequential methods, homophily is by far the strongest for political affiliation:
\rr{observed cross-party relations are 82\% less frequent than expected under Local and 66\% less frequent than expected under Sequential.
Rates of political homophily are even stronger for the other LLMs that we test, most extremely for GPT-4o and Llama 3.1 70B, where none of the edges are cross-party and the network fractures into two disconnected components (Appendix~\ref{sec:diff-llms}).}

Does this mean that LLMs actually pay the most attention to political affiliation when choosing social ties?
Despite political homophily being the strongest, this could be due to correlations between political affiliation and other demographics; for example, hypothetically, if all Democrats had the same gender and race, and all Republicans had the same gender and race, then apparent homophily in political affiliation could actually be due to similarity in other demographics.
To test this, we try shuffling the demographics, so that, while maintaining the same numbers of each group per demographic, each persona is randomly assigned to a group, thus removing correlations between demographics.
When we run Sequential with these shuffled personas, we find that political homophily remains by far the strongest (Figure~\ref{fig:homophily}), demonstrating that the LLM is, in fact, paying most attention to political affiliation when choosing social ties.
\rr{As an additional test, we also try ablations of the demographics, where we present the LLM with only one demographic variable at a time, or with two variables, one being political affiliation and the other being one of the four others. Here, we also find that political affiliation continues to be the dominant factor: when only one variable is presented, levels of homophily increase for \textit{all} demographics but it remains the highest for political affiliation; when two variables are presented, political homophily is always stronger than the other demographic's homophily (Figure \ref{fig:ablations}).}

\begin{table}[t]
    \small
    \centering
    \begin{tabular}{c|c}
        \textbf{Demographic} & \textbf{Reason} \% \\
        \hline 
        Political affiliation & 86.7\% \\
        Religion & 43.0\% \\
        Age & 21.8\% \\
        Race/ethnicity & 12.1\% \\
        Gender & 7.3\%
    \end{tabular}
    \caption{Frequency that each demographic is part of the LLM's reason for choosing a persona as a friend.}
    \label{tab:reasons}
\end{table}
\begin{table*}[t]
    \centering
    \small
    \begin{tabular}{p{2.6cm}||p{1cm}|p{1.7cm}||p{3.8cm}|p{3.2cm}|p{1.6cm}}
        \textbf{Measure} & \textbf{Local} & \textbf{Sequential} & \textbf{Source} & \textbf{Description} & \textbf{Value} \\
        \hline 
        \rr{Cross-group ratio (Eq.~\ref{eqn:cross-group-ratio}) $\downarrow$} & \rr{0.180 (0.015)} & \rr{0.340 (0.022)} & \rr{\citet{halberstam2016homophily}} & \rr{Twitter} & \rr{0.528} \\
        \hline 
        Same-group ratio (Eq.~\ref{eqn:same-group-ratio}) $\uparrow$ & 1.851 (0.016) & 1.685 (0.023) & \citet{halberstam2016homophily} & Twitter & 1.404 \\
        \hline 
        Isolation index (Eq.~\ref{eqn:isolation}) $\uparrow$ & 0.729 (0.020) & 0.530 (0.027) & \citet{halberstam2016homophily} & Twitter & 0.403 \\
         & & & \citet{gentzkow2011segregation} & Voluntary associations & 0.145 \\
        & & & & Work & 0.168 \\ 
        & & & & Neighborhood & 0.187 \\ 
        & & & & Family & 0.243 \\
        & & & & People you trust & 0.303 \\
        & & & & Political discussants & 0.394 \\
        \hline 
        Polarization (Eq.~\ref{eqn:polarization}) $\uparrow$ & 0.639 (0.037) & 0.515 (0.041) & \citet{garimella2017twitter} & Twitter, follow & 0.33-0.42 \\
        & & & & Twitter, retweet & 0.37-0.41 \\
    \end{tabular}
    \caption{Comparing political homophily in our generated networks to real-world networks. \rr{We consider different measures of homophily and indicate with $\downarrow$ or $\uparrow$ the direction that indicates greater homophily for that measure.} For Local and Sequential, we report the mean and standard error (in parentheses) over each method's 30 generated networks.}
    \label{tab:political}
\end{table*}
Finally, we directly test what the LLM is paying attention to by prompting it to generate a short reason for each friend that it selects.
Then, given the reason, we use GPT-4o to classify the reason, e.g., ``I'm a woman too and we share the same religion and political affiliation'' is classified as [gender, religion, political affiliation].
We allow the LLM to generate free-text reasons during network generation since we do not want to constrain its response (e.g., if it is using other information, such as degree or ID).
We find that political affiliation strongly dominates here as well: it is part of the reason 86.7\% of the time, while the next most-mentioned demographic, religion, is only mentioned 43.0\% of the time (Table~\ref{tab:reasons}). 
As a caveat, prior work has shown that we cannot always trust an LLM's own explanation for its choices \cite{agarwal2024faithfulness}.
However, given the alignment of these results with our other results, there is strong evidence that LLMs pay the most attention to political affiliation when generating social relations.

\paragraph{LLMs \textit{overestimate} political homophily.}
Given the LLM's emphasis on political homophily, we seek to compare its level of political homophily to reported levels from real-world social networks.
We are not able to compare to the eight social networks from Section~\ref{sec:structure} here since we do not have demographic features per node.
However, we are able to find reported political homophily in several papers, covering both online and offline social networks.
In Table~\ref{tab:political}, we summarize these comparisons, showing that Local and Sequential consistently overestimate political homophily across different measures of homophily. 
For example, \citet{halberstam2016homophily} analyze political homophily on Twitter.
\rr{In their data, cross-party relations appear 47\% less often than expected, which indicates homophily, but not as strong as what the LLM predicts.
In addition to the cross-group ratio, we can also compute a \textit{same-group} ratio, using the ratio of observed-to-expected proportion of same-group edges, where a higher ratio indicates more homophily (Eq. \ref{eqn:same-group-ratio}).
Using this measure, we find that same-party relations only appear 40\% more often than expected in the Twitter data, while same-party relations are 85\% and 68\% more frequent than expected for Local and Sequential, respectively.}

\citet{halberstam2016homophily} also compute the isolation index, which is the difference in average conservative exposure between conservatives and liberals (Eq. \ref{eqn:isolation}), with larger indices indicating greater isolation.
They find an isolation index of 0.403, while the LLM's is far higher, at 0.720 for Local and 0.530 for Sequential.
Furthermore, the authors note that homophily could be overestimated in their data, since they selected users who follow politicians, which ``may tend to disproportionately include individuals with strong preferences for linking to like-minded users.''
Even though their homophily results could be overestimates, the LLM's estimates still significantly exceed theirs. 

Isolation indices are even lower in \citet{gentzkow2011segregation}, who study ideological segregation in social networks.
In face-to-face interactions, the highest isolation indices they report are 0.243 (family), 0.303 (people you trust) and 0.394 (among people who discuss politics).
Finally, we also compare to \citet{garimella2017twitter}, who study political polarization on Twitter and define it as $p_i = 2 \cdot |0.5 - \alpha / (\alpha + \beta)|$, where $\alpha$ and $\beta$ indicate how many left-leaning and right-learning users, respectively, are followed by user $i$, and they take the average $p_i$ over users (Eq. \ref{eqn:polarization}).
Their measure captures the difference between observed leaning and a balanced leaning of $0.5$, with higher numbers indicating greater polarization.
They report polarization levels of 0.33-0.42 for following relations and 0.37-0.41 for retweets (increasing over time), while we find significantly higher levels in the LLM's generated networks, with 0.639 for Local and 0.515 for Sequential.

We hypothesize that the LLM overestimates political homophily partially due to high levels of polarization in its online pretraining data and frequent discussions of such polarization, although future work is needed to carefully study this phenomena.
These results also have important implications if one seeks to run experiments over social networks generated by LLMs.
For example, overestimated political homophily may result in unrealistically high levels of polarization, 
potentially leading to incorrect conclusions and interventions.

\begin{figure*}[t!]
    \centering
    \includegraphics[width=\linewidth]{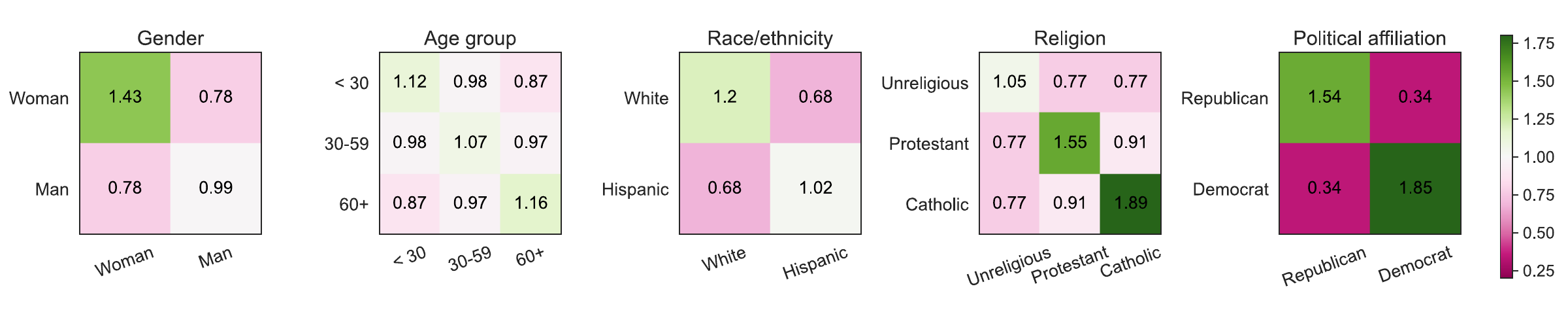}
    \caption{Ratio of observed-to-expected proportion of edges, for all pairs of demographic groups (Eq.~\ref{eqn:pair-ratio}). All subfigures share the same colormap (right). Groups with at least 10 nodes are kept.}
    \label{fig:pairwise-ratios}
\end{figure*}

{\paragraph{Homophily between pairs of groups.}
\rr{
As described in Eq. \ref{eqn:cross-group-ratio}, we measure homophily as the ratio of observed-to-expected proportion of cross-group edges.
Now, we extend this definition to compute an observed-to-expected ratio for any pair of groups, $A$ and $B$ (where $A$ could equal $B$):
\begin{align}
    H_{AB} &= \frac{\frac{\sum_{ij} A_{ij} \cdot \mathds{1}[g_i = A] \cdot \mathds{1}[g_j = B]}{E}}{\frac{N_A (N_B - \mathds{1}[A = B])}{N(N-1)}}. \label{eqn:pair-ratio}
\end{align}
This measures the observed proportion of edges that are between nodes in groups $A$ and $B$ divided by the expected proportion of edges between nodes in these groups. 
In Figure~\ref{fig:pairwise-ratios}, we visualize $H_{AB}$ for all pairs, for each of the five demographic variables.
For each variable, we keep all groups with at least 10 personas in our set of 50 personas (Table~\ref{tab:personas}), and compute $H_{AB}$ for each pair of groups, reporting the mean ratio over the 30 generated networks from Sequential.}

\rr{We find, as expected, that the diagonal (i.e., same-group ratios) tends to be above 1, although notably it is not for relations between men, with a ratio of 0.99, while relations between women have a ratio of 1.43. 
Thus, even within one demographic variable, the LLM's levels of homophily vary for different groups, such as homophily within women being stronger than homophily within men.
Variability across groups is present for all demographic variables: most extremely, for religion, relations between Catholics occur 89\% more frequently than expected, while relations are only 5\% more frequent for personas that identify as Unreligious.
This plot also reveals that not all cross-group relations are equally unlikely: for example, when we divide age into three age groups, the adjacent age groups have higher cross-group ratios (0.98 for under 30 and 30-59 and 0.97 for 30-59 and 60+) compared to the non-adjacent age groups (0.87 for under 30 and 60+).
Finally, this plot reiterates that political affiliation has the lowest cross-group ratio, at 0.34, and reveals that it also has among the highest same-group ratios, although the LLM gives Democrats stronger same-group preferences, with a same-group ratio of 1.85, than Republicans, with a same-group ratio of 1.54.}

\subsection{Incorporating interests}
\label{sec:interests}
\begin{table*}[]
    \small
    \centering
    \begin{tabular}{p{1.1cm}|p{2.4cm}|p{0.6cm}|p{1.7cm}|p{1.8cm}|p{6.8cm}}
         \textbf{Gender} & \textbf{Race / Ethnicity} & \textbf{Age} & \textbf{Religion} & \textbf{Political Affiliation} & \textbf{Interests}\\
         \hline 
         Man & White & 47 & Protestant & Republican & Hunting, fishing, classic rock, church activities, patriotic events, home improvement \\
         Woman & Black & 69 & Unreligious & Republican & History, gardening, community service, classic jazz music, financial news, travel \\
         Man & Hispanic & 75 & Unreligious & Democrat & Historical documentaries, community events, family gatherings, literature on social justice \\
         Man & American Indian / Alaska Native & 30 & Protestant & Republican & Outdoor activities, traditional crafts, conservative politics, music, community service, history\\
         Woman & Asian & 58 & Catholic & Democrat & Volunteering, social justice, culinary arts, family activities, church community involvement \\
    \end{tabular}
    \caption{Examples of LLM-generated interests for personas with different demographics.}
    \label{tab:interests}
\end{table*}
A natural question is whether demographic homophily is exaggerated because we only give the LLM demographic information, without other important details such as the person's interests.
Thus, we run an additional set of experiments where we allow the LLM to also generate interests for each persona.
To generate interests, we prompt the LLM
with, ``In 8-12 words, describe the interests of someone with the following demographics'' (full prompt in Figure~\ref{fig:interest-prompt}).
In Table \ref{tab:interests}, we provide examples of the generated interests, with the full list of personas with interests available online.

\paragraph{Effect of interests on networks.}
As shown in Figure~\ref{fig:homophily}, by comparing Sequential vs. Sequential with interests, we find that incorporating interests strengthens homophily for some demographics (clearly for gender \rr{and age} and slightly for political affiliation) and weakens homophily for the others (race/ethnicity and religion).
Notably, after these adjustments, political homophily remains the strongest across the demographic variables.
Thus, our finding that LLMs prioritize political homophily is robust to incorporating interests.

In fact, the interests themselves strongly encode political homophily, with evidence of political stereotyping.
For example, among the Democrat personas, the most common interests are ``social justice'' (62.5\% of Democrats vs. 0.2\% of Republicans), ``community service'' (29.3\% vs. 13.9\%), and ``progressive policies'' (18.6\% vs. 0\%).
In contrast, among the Republican personas, the most common interests are ``conservative politics'' (41.6\% of Republicans vs. 0\% of Democrats), ``church activities'' (32.1\% vs. 13.2\%), and ``gardening'' (23.2\% vs 16.4\%). 
We provide the complete list of top 10 interests per demographic group in Table~\ref{tab:top-interests-per-demo}.
\begin{figure}
    \centering
    \includegraphics[width=\linewidth]{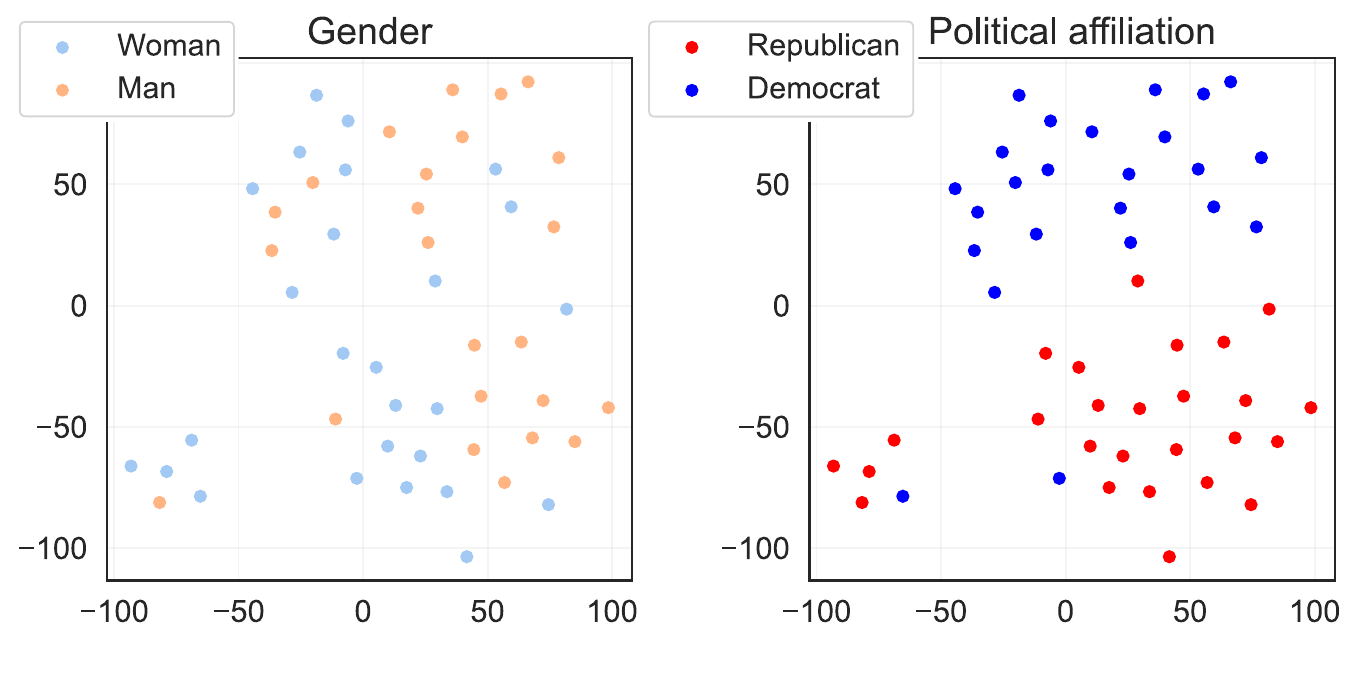}
    \caption{Embeddings of interests, with T-SNE projection to 2D. Each dot is a persona, colored by its gender (left) or its political affiliation (right).}
    \label{fig:interest-embs}
\end{figure}
We also analyze interests by mapping them to text embeddings, using OpenAI's text-embedding-3-small model.
In Figure~\ref{fig:interest-embs}, we visualize the embeddings, coloring them by their persona's gender (left) and political affiliation (right).
We compare these two demographic variables, since both have two, approximately equally sized groups in our sample of 50 personas.
From this comparison, we can see how much more distinct the political groups are than the gender groups, demonstrating the level of political homophily encoded in the interests.
Finally, we conduct a supplementary experiment where we try network generation with \textit{only} interests and no demographics.
Homophily decreases across all demographic variables, but political homophily remains by far the strongest (Figure~\ref{fig:only-interests}).

These results demonstrate how the LLM's emphasis on political affiliation also appears in interest generation, and, as a result, incorporating LLM-generated interests cannot help to reduce overestimated political homophily. 
These results build on prior results showing that LLM sometimes exhibit bias in political settings \cite{cheng2023compost,santurkar2023opinions,wang2024cannot}, exploring these issues through a novel lens of homophily and social network generation. 
\section{Discussion}
\label{sec:discussion}
Our work has established several findings.
First, with the right prompting method, the LLM is able to simultaneously match many structural characteristics of real social networks, outperforming classical network models \rr{with few parameters}.
Second, ``local'' prompting methods produce more realistic networks than ``global'' methods and, within local methods, adding network information (i.e., Sequential) helps the LLM capture long-tailed degree distributions.
Third, the LLM exhibits clear homophily across five key demographic variables, but political homophily dominates, to the extent that it is overestimated relative to real-world measures.
Finally, incorporating LLM-generated interests does not reduce these overestimates, since the interests themselves encode strong political homophily.

\paragraph{Future directions.}
Our findings demonstrate the promise of generating social networks with LLMs, as they are zero-shot, flexible, and structurally realistic.
However, more needs to be done to address potential biases, especially with regards to political homophily.
\rr{One possibility could be to incorporate more information per persona: we found in our ablations that the LLM tends to produce stronger homophily per variable when fewer variables are provided (Figure \ref{fig:ablations}), so we might see political homophily reduce if more variables or richer descriptions (such as interests) are provided per personas.
However, given our findings on LLM-generated interests, researchers may want to handcraft interests with fewer political stereotypes, although other methods would be needed to scale such efforts to large social networks.
Furthermore, while we focus in this work on how LLMs overestimate political homophily compared to real-world measures, in the Appendix we discuss real-world measures of homophily for the other demographic variables and find that the LLM, in fact, seems to \textit{underestimate} homophily for race/ethnicity and religion (Table \ref{tab:homophily-other-demos}), calling for future work that thoroughly investigates potential biases along the other demographic dimensions.}

There are also other limitations to LLM social network generation: for example, while we have shown that Sequential can approximately match the \textit{means} of the real networks' metrics, its generated networks consistently demonstrate less \textit{variance} than the real networks (Figure~\ref{fig:metrics}), reflecting known issues of LLMs to flatten demographic groups \cite{wang2024cannot} and lack output diversity \cite{kirk2023diversity}.
Furthermore, all of our methods require listing $N$ personas per prompt, which becomes infeasible with larger networks, due to context windows and cost.
In Table~\ref{tab:costs}, we conduct a big-O analysis of how the number of tokens scales with network size, revealing a tradeoff between the improved realism of Sequential and Local versus the lower costs of Global.
\rr{To address this tradeoff, future work could explore how to make the Sequential or Local methods more scalable: for example, instead of listing $N$ personas per prompt, where each persona chooses from the full list of all other personas, one could provide a \textit{subset} of the other personas. In Appendix \ref{sec:app-network}, we discuss a simple implementation of this idea where the subsets are sampled uniformly at random, and show how this extension enables us to generate networks with 300 personas (sampling 30 personas per prompt). These larger networks are similar to our smaller networks in terms of homophily and some structural characteristics, although density and clustering drop due to the subsampling (Figure \ref{fig:networks-300}).}

In future work, we hope to \rr{further extend these methods to make them both scalable and realistic,} as well as generate networks with greater variance.
We also hope to study whether our results on political homophily generalize outside of the US, and to explore methods to mitigate the political biases we observe in this work.

\section*{Acknowledgments}
We thank Joon Sung Park and Marios Papachristou for helpful comments.
S.C. was supported in part by a Meta PhD Fellowship and NSF Graduate Research Fellowship.
We also gratefully acknowledge the support of
NSF under Nos. OAC-1835598 (CINES), CCF-1918940 (Expeditions), DMS-2327709 (IHBEM);
Stanford Data Applications Initiative,
Wu Tsai Neurosciences Institute,
Stanford Institute for Human-Centered AI,
Chan Zuckerberg Initiative,
Amazon, Genentech, GSK, Hitachi, SAP, and UCB.

\bibliography{references}
\section*{Paper Checklist}

\begin{enumerate}

\item For most authors...
\begin{enumerate}
    \item  Would answering this research question advance science without violating social contracts, such as violating privacy norms, perpetuating unfair profiling, exacerbating the socio-economic divide, or implying disrespect to societies or cultures?
    \answerYes{Yes, our work generates synthetic networks over hypothetical individuals, so it does not violate privacy norms, and we specifically analyze potential social harms, such as stereotyping or exacerbating segregation.}
  \item Do your main claims in the abstract and introduction accurately reflect the paper's contributions and scope?
    \answerYes{Yes, the claims in the abstract and introduction are supported by our main results in Section~\ref{sec:results}, which are further tested with robustness checks in our Appendices.}
   \item Do you clarify how the proposed methodological approach is appropriate for the claims made? 
    \answerYes{Yes, to analyze LLMs' social network generation, we need to both provide methods for generating social networks with LLMs (Section~\ref{sec:prompting-methods}) and an evaluation framework (Section~\ref{sec:metrics}).}
   \item Do you clarify what are possible artifacts in the data used, given population-specific distributions?
    \answerYes{Yes, we specify how we sample persona demographics following the US distribution (Section~\ref{sec:prompting-methods} and Appendix~\ref{sec:persona-details}), and we describe which populations are captured in the real social networks (Section~\ref{sec:metrics} and Appendix~\ref{sec:app-structure}). The rest of the data is generated by LLMs, i.e., the generated social networks, which are our object of study, so we study their artifacts closely.}
  \item Did you describe the limitations of your work?
    \answerYes{Yes, we discuss limitations in the Discussion (Section~\ref{sec:discussion}), with additional results and discussion referred to in the Appendices.}
   \item Did you discuss any potential negative societal impacts of your work?
    \answerYes{Yes, we discuss potential negative impacts of generating social networks with LLMs, particularly related to bias and stereotyping. We discuss related work on bias in social settings (Section~\ref{sec:related}) and investigate such biases in our work, documenting political biases in the generated networks and generated interests (Section~\ref{sec:results}).}
    \item Did you discuss any potential misuse of your work?
    \answerYes{Yes, we point out that it would be risky to simulate politically-related social phenomena, such as polarization or opinion dynamics, over networks generated by LLMs, since we find that they overestimate political homophily.}
    \item Did you describe steps taken to prevent or mitigate potential negative outcomes of the research, such as data and model documentation, data anonymization, responsible release, access control, and the reproducibility of findings?
    \answerYes{Yes, we have carefully documented the risks of generating social networks with LLMs, such as overestimated political homophily, underestimated variance, and LLM costs. Our code, data, and generated networks are available at \github. This includes all code to run our experiments and reproduce our results.}
  \item Have you read the ethics review guidelines and ensured that your paper conforms to them?
    \answerYes{Yes, we have.}
\end{enumerate}

\item Additionally, if your study involves hypotheses testing...
\begin{enumerate}
  \item Did you clearly state the assumptions underlying all theoretical results?
    \answerYes{NA, we do not have theoretical results.}
  \item Have you provided justifications for all theoretical results?
    \answerYes{Yes, we have provided evidence to support all claims, see Section~\ref{sec:results} and Appendices.}
  \item Did you discuss competing hypotheses or theories that might challenge or complement your theoretical results?
    \answerYes{Yes. For example, we test whether the fact that political homophily is the highest can be explained by correlations between demographics, so we try shuffling demographics, and we find that political homophily is still the highest. We also test six different LLMs, to test how broadly our results generalize. Finally, we test minor changes to the experimental set-up (e.g., temperature and prompt), and we show that our main results are not sensitive to these changes.}
  \item Have you considered alternative mechanisms or explanations that might account for the same outcomes observed in your study?
    \answerYes{Yes, see above.}
  \item Did you address potential biases or limitations in your theoretical framework?
    \answerNA{NA, we do not have a theoretical framework.}
  \item Have you related your theoretical results to the existing literature in social science?
    \answerYes{Yes, we discuss related work from sociology \cite{watts1998collective,krackhardt1999ties,mcpherson2001birds,kossinets2009homophily,smith2014gss}, sociometry \cite{coleman1957diffusion,macrae1960sociometric}, economics \cite{gentzkow2011segregation,halberstam2016homophily}, and political science \cite{argyle2023samples}.}
  \item Did you discuss the implications of your theoretical results for policy, practice, or further research in the social science domain?
    \answerYes{Yes, we discuss the implications in Section~\ref{sec:discussion} and the Appendices.}
\end{enumerate}

\item Additionally, if you are including theoretical proofs...
\begin{enumerate}
  \item Did you state the full set of assumptions of all theoretical results?
    \answerNA{NA, we do not have theoretical proofs.}
   \item Did you include complete proofs of all theoretical results?
    \answerNA{NA, we do not have theoretical proofs.}
\end{enumerate}

\item Additionally, if you ran machine learning experiments...
\begin{enumerate}
  \item Did you include the code, data, and instructions needed to reproduce the main experimental results (either in the supplemental material or as a URL)?
    \answerYes{Yes, all code, data, and instructions are available at \github, which is included on the first page.}
  \item Did you specify all the training details (e.g., data splits, hyperparameters, how they were chosen)?
    \answerYes{We do not train models, but we do describe experimental parameters, such as LLM temperature, the number of personas sampled, the number of networks sampled, randomization in the prompts, etc.}
     \item Did you report error bars (e.g., with respect to the random seed after running experiments multiple times)?
    \answerYes{Yes, we generate 30 networks per prompting method and report mean, standard error, and individual data points.}
    \item Did you include the total amount of compute and the type of resources used (e.g., type of GPUs, internal cluster, or cloud provider)?
    \answerYes{Yes, we describe the APIs we used---OpenAI API and Llama API---that we used to call the models. In our GitHub repository, we also describe our version of Python (3.10) and list the required Python packages and versions.}
     \item Do you justify how the proposed evaluation is sufficient and appropriate to the claims made? 
    \answerYes{Yes, in Section~\ref{sec:metrics}, we describe our evaluation framework, including the real networks that we compare against and the network characteristics that we measure.}
     \item Do you discuss what is ``the cost`` of misclassification and fault (in)tolerance?
    \answerNA{NA, we do not have a classification problem.}
  
\end{enumerate}

\item Additionally, if you are using existing assets (e.g., code, data, models) or curating/releasing new assets,
\begin{enumerate}
  \item If your work uses existing assets, did you cite the creators?
    \answerYes{Yes, we cited our data sources, which included statistics to sample demographics following the US population and the real social networks.}
  \item Did you mention the license of the assets?
    \answerNA{NA, these data sources are all publicly available.}
  \item Did you include any new assets in the supplemental material or as a URL?
    \answerYes{Yes, we provide our code (with methods to generate social networks) and generated networks in our GitHub repository.}
  \item Did you discuss whether and how consent was obtained from people whose data you're using/curating?
    \answerNA{NA, we are generating synthetic networks over hypothetical individuals.}
  \item Did you discuss whether the data you are using/curating contains personally identifiable information or offensive content?
    \answerYes{Yes, we describe that the real networks do not contain node features, so they are not personally identifiable, and the population statistics are aggregated over the entire US population.}
    \item If you are curating or releasing new datasets, did you discuss how you intend to make your datasets FAIR (see \citet{fair})?
    \answerNA{NA, we are not releasing new datasets.}
    \item If you are curating or releasing new datasets, did you create a Datasheet for the Dataset (see \citet{gebru2021datasheets})? 
    \answerNA{NA, we are not releasing new datasets.}
\end{enumerate}

\item Additionally, if you used crowdsourcing or conducted research with human subjects, 
\begin{enumerate}
  \item Did you include the full text of instructions given to participants and screenshots?
    \answerNA{NA, our work does not involve crowdsourcing or research with human subjects.}
  \item Did you describe any potential participant risks, with mentions of Institutional Review Board (IRB) approvals?
    \answerNA{NA, our work does not involve crowdsourcing or research with human subjects.}
  \item Did you include the estimated hourly wage paid to participants and the total amount spent on participant compensation?
    \answerNA{NA, our work does not involve crowdsourcing or research with human subjects.}
   \item Did you discuss how data is stored, shared, and deidentified?
   \answerNA{NA, our work does not involve crowdsourcing or research with human subjects.}
\end{enumerate}

\end{enumerate}

\appendix
\section*{Appendix}
In Appendix~\ref{sec:diff-llms}, we compare results across GPT, Llama, and Gemma models.
In Appendix~\ref{sec:app-results}, we describe additional experiments and findings. 
In Appendix~\ref{sec:app-methods}, we provide details on methods and experimental set-up.

\renewcommand{\thefigure}{\thesection\arabic{figure}}
\renewcommand{\thetable}{\thesection\arabic{table}}
\setcounter{figure}{0}
\setcounter{table}{0}
\section{Comparison Between LLMs}
\label{sec:diff-llms}
\begin{figure*}
    \centering
    \includegraphics[width=\linewidth]{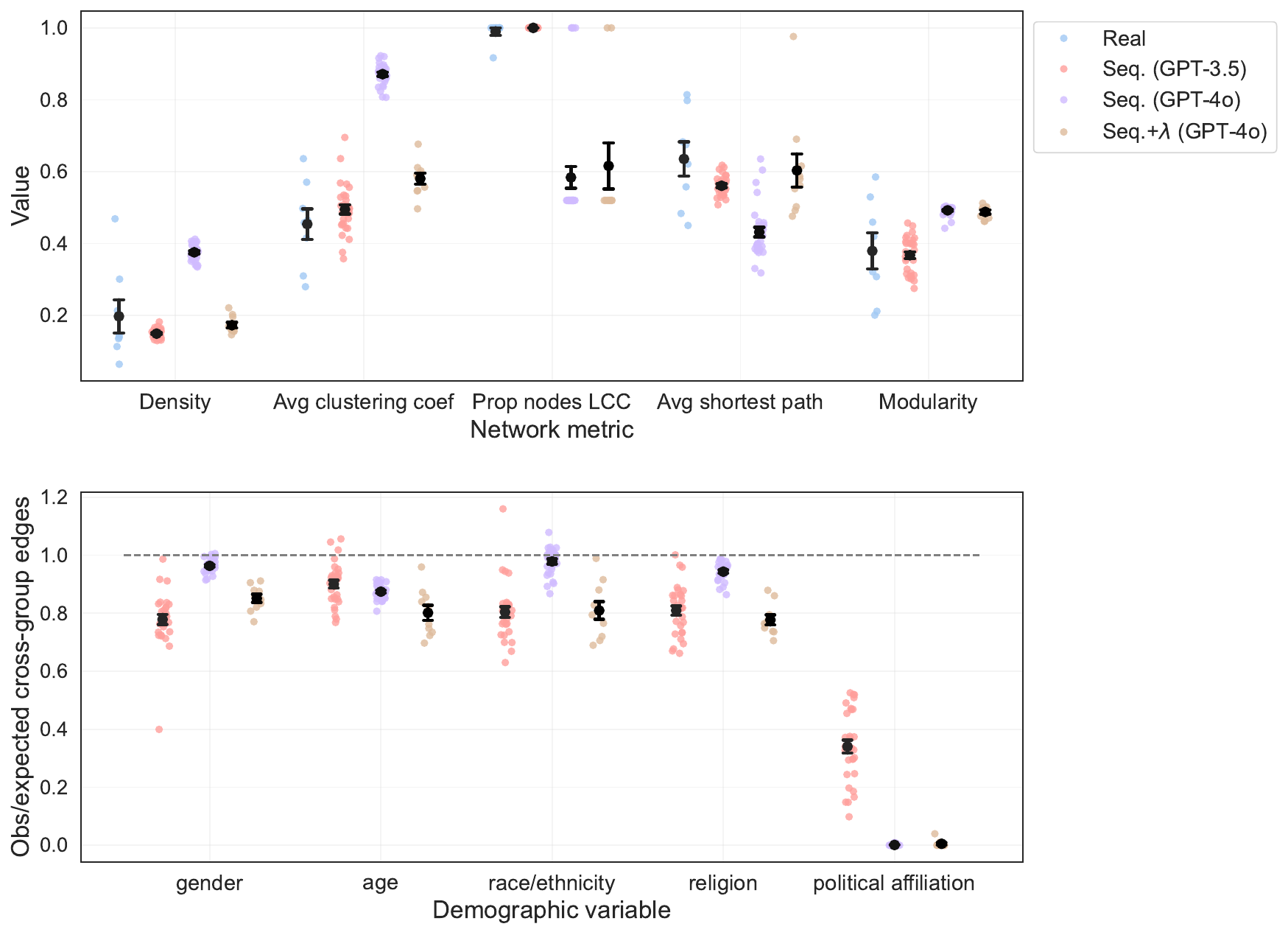}
    \caption{
    Evaluating results from GPT-4o, compared to GPT-3.5 Turbo.
    \textbf{Top:} structural network metrics. \textbf{Bottom:} homophily, where ratios \rr{below} 1 (marked by the grey line) indicate homophily and \rr{lower} ratios indicate more homophily.
    In both plots, we visualize mean and standard error (in black) and individual data points corresponding to each network.
    }
    \label{fig:gpt4o}
\end{figure*}

\begin{figure*}
    \centering
    \includegraphics[width=\linewidth]{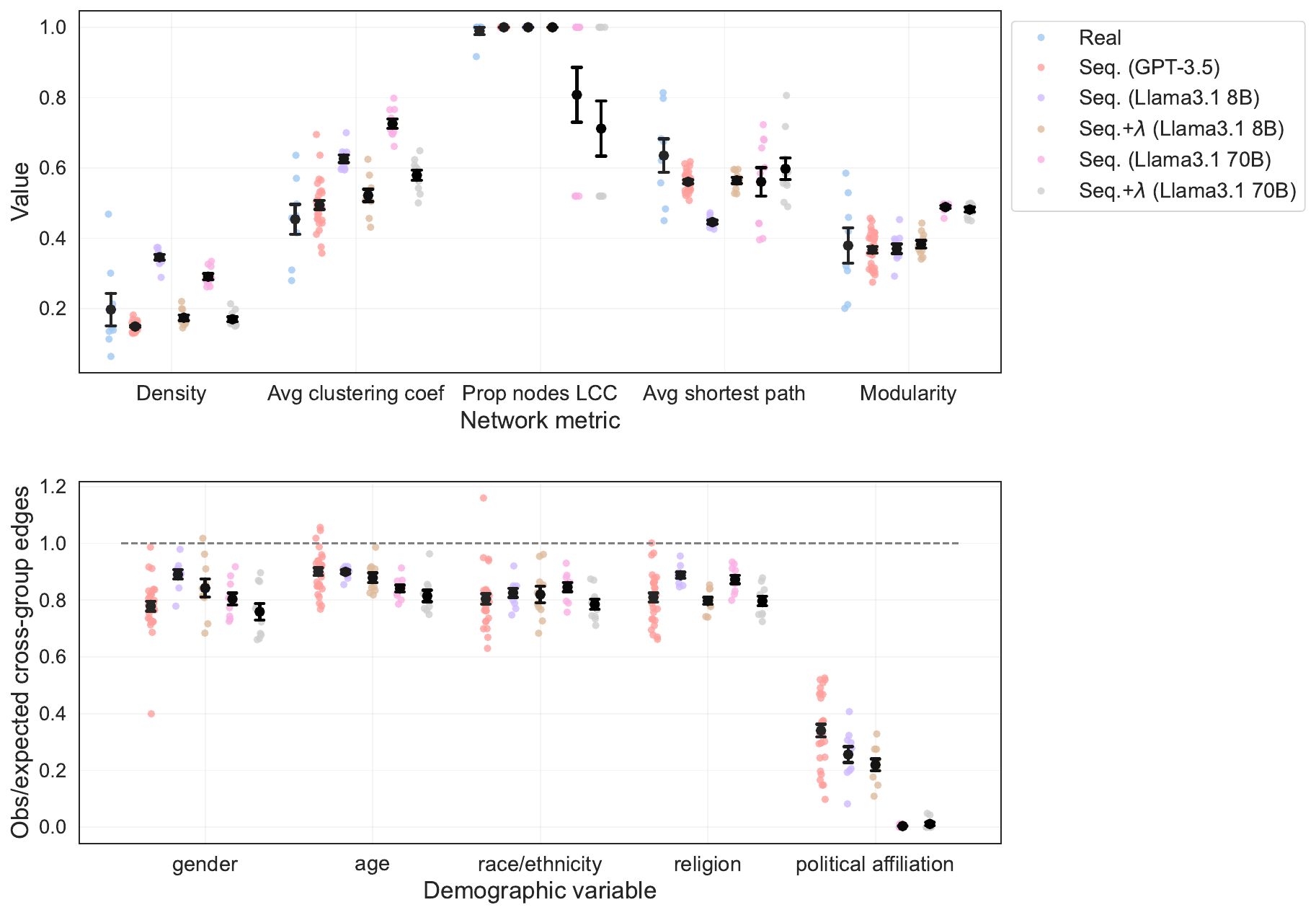}
    \caption{
    Evaluating results from Llama 3.1 8B and 70B, compared to GPT-3.5 Turbo.
    \textbf{Top:} structural network metrics. \textbf{Bottom:} homophily, where ratios \rr{below} 1 (marked by the grey line) indicate homophily and \rr{lower} ratios indicate more homophily.
    In both plots, we visualize mean and standard error (in black) and individual data points corresponding to each network.
    }
    \label{fig:llama}
\end{figure*}

\begin{figure*}
    \centering
    \includegraphics[width=\linewidth]{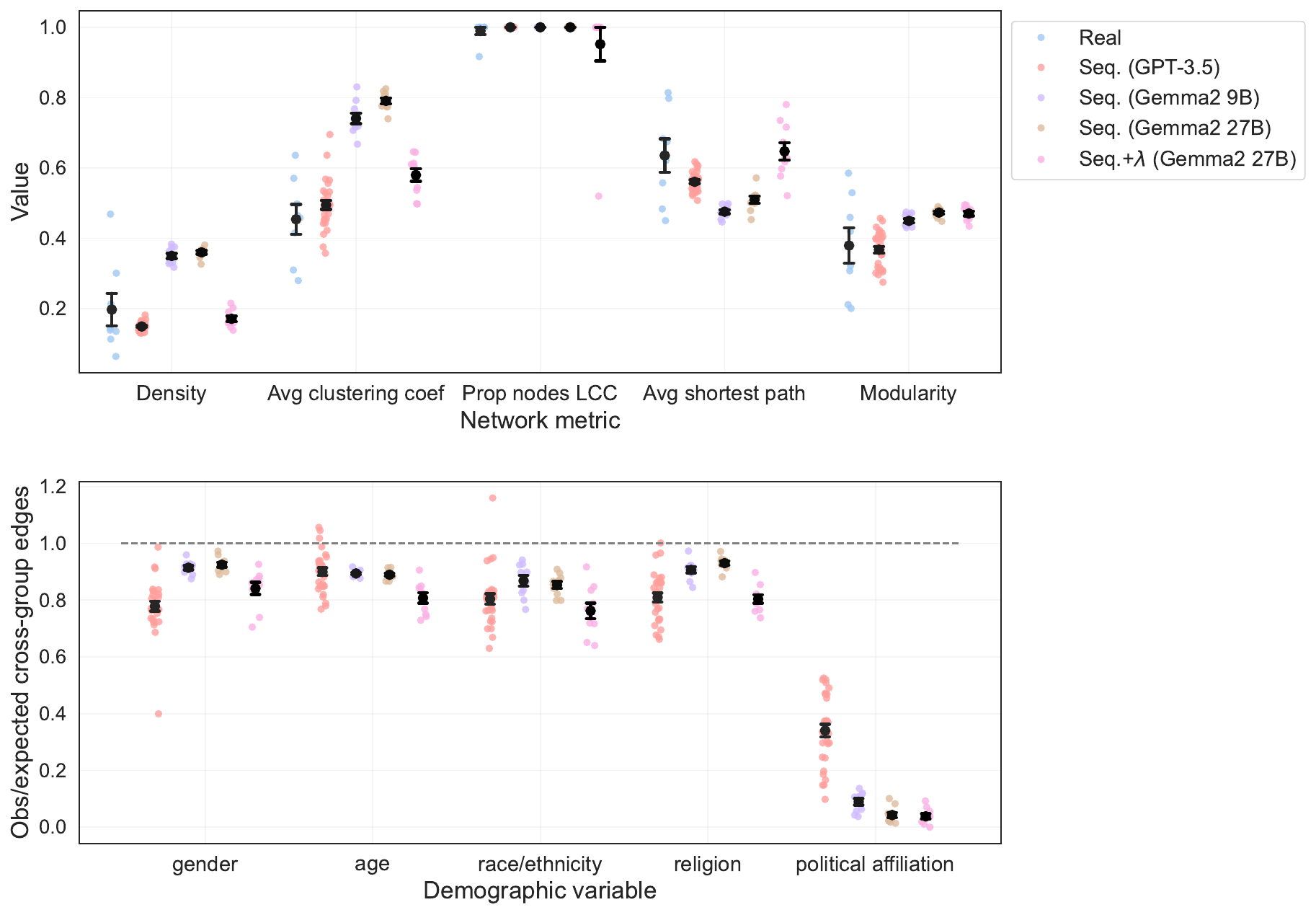}
    \caption{
    Evaluating results from Gemma 2 9B and 27B, compared to GPT-3.5 Turbo.
    \textbf{Top:} structural network metrics. \textbf{Bottom:} homophily, where ratios \rr{below} 1 (marked by the grey line) indicate homophily and \rr{lower} ratios indicate more homophily.
    In both plots, we visualize mean and standard error (in black) and individual data points corresponding to each network.
    }
    \label{fig:gemma}
\end{figure*}

We experiment with the following LLMs: OpenAI's GPT-3.5 Turbo and GPT-4o \cite{brown2020gpt3,openai2023gpt4}, Meta's Llama 3.1 (8B and 70B) \cite{llama2023}, and Google's Gemma 2 (9B and 27B) \cite{gemma2024}.
\rr{These six models represent a range across companies, different model sizes, and proprietary (the GPT models) versus open-source (the Llama and Gemma models). Llama 3.1 in particular includes some of the best-performing open-source models on holistic benchmarks.}\footnote{\rr{\url{https://crfm.stanford.edu/helm/lite/latest/#/leaderboard}}}.
For the OpenAI models, we use the OpenAI API,\footnote{\url{https://platform.openai.com/docs/api-reference}} and for the other models, we use the Llama API,\footnote{\url{https://www.llama-api.com/}} which also includes other open-source models.
We report our main results on GPT-3.5 Turbo since we find that it best matches the structure of the real-world social networks (discussed below), but here, we discuss results from all other models.
For these experiments, we generate networks with the Sequential method, using the same experimental settings as before (same set of 50 personas, prompt, temperature of 0.8, etc.).
For these experiments, we generate 10 instead of 30 networks per model, but we find that standard errors are small.
We visualize results for GPT-4o in Figure~\ref{fig:gpt4o}, Llama 3.1 8B and 70B in Figure~\ref{fig:llama}, and Gemma 2 9B and 27B in Figure~\ref{fig:gemma}.

\paragraph{Structural characteristics.}
We find that GPT-3.5 Turbo best matches the structure of the real networks, most notably matching the real-world density. 
All of the other models have much higher densities, which also contributes to unrealistically high clustering and low shortest paths.
Due to density's outsized effect, we wanted to see if providing the model a bit of help on density might be all that it needs to match the other characteristics as well.
Thus, we try a variant of the Sequential method where we specify $n$, i.e., how many friends should be chosen (instead of only asking ``Which of these people will you become friends with?'', see full prompt in Figure~\ref{fig:sequential-prompt}).
We sample $n$ from Exponential($\lambda=0.2$), with mean $1/\lambda=5$, independently for each persona.
Note that specifying each persona's number of choices does not predetermine the exact density or degree distribution of the network, since a persona's total set of friends at the end of the network generating process is the union of its chosen friends along with anyone who chose it.
However, specifying these numbers can help to guide the models to lower densities.

With this variant, which we call ``$+\lambda$'', all models' densities are brought down to a reasonable range.
Llama 3.1 8B $+\lambda$ is also able to approximately match the real networks and GPT-3.5 Turbo on all other characteristics now (Figure~\ref{fig:llama}).
However, even with $+\lambda$, GPT-4o and Llama 3.1 70B still generate networks that are often disconnected into two components; this is due to extreme political segregation, which we discuss below.
With Gemma 2 27B, the networks are almost always fully connected, but even with $+\lambda$, clustering and modularity remain slightly too high.\footnote{We were not able to add $+\lambda$ to Gemma 2 9B, since it could not consistently follow instructions to choose exactly $n$ friends.}

\paragraph{Homophily.}
We find that all six models exhibit clear homophily, with ratios above 1 for gender, age, race/ethnicity, religion, and political affiliation.
We also find, consistently across the models, that political homophily remains by far the strongest form of homophily and it is always overestimated compared to real-world measures.
In fact, the models we test in this section all produce levels of political homophily even higher than GPT-3.5 Turbo, which we discussed were overestimates in the main text (Table~\ref{tab:political}).

The emphasis on political homophily is particularly extreme for GPT-4o and Llama 3.1 70B: they show complete segregation between Republicans and Democrats, such that there are \textit{no} cross-group edges and the cross-group ratio is around 0 (Figures~\ref{fig:gpt4o} and \ref{fig:llama}).
Furthermore, as a result of the complete segregation, their networks fracture into two disconnected components, one for Republicans and one for Democrats, which is why we see that the proportion of nodes in the largest connected component is often around 50\% for these two models.
This is highly unrealistic, as it is well-known that social networks, despite having strong community structure and homophily, are also characterized by having a giant connected component that contains the vast majority of the nodes \cite{ugander2011facebook}.
Splitting a synthetic network into two disconnected components has important implications for downstream modeling and use cases of these generated networks: for example, an epidemic outbreak started in one component would never reach the other component, or opinions could not spread, greatly altering the trajectory of dynamic network processes.
\setcounter{figure}{0}
\setcounter{table}{0}
\section{Additional Results}
\label{sec:app-results}
\begin{table*}[t]
    \centering
    \begin{tabular}{c||c|c|c|c|c|c||c}
        \textbf{Model} & \textbf{Density} & \textbf{Avg CC} & \textbf{\% LCC} & \textbf{Avg SP} & \textbf{Mod.} & \textbf{Degree} & \textbf{Avg} \\
        \hline 
        Random graph & 0.013$^\dagger$ & 2.286 & \textbf{0.378} & 1.156 & 1.149 & 0.154 & 1.025 \\
        Preferential attachment & 0.113$^\dagger$ & 1.363 & \textbf{0.378} & 1.081 & 1.160 & 0.053 & 0.807 \\
        Small world & 0.054$^\dagger$ & 0.041$^\dagger$ & \textbf{0.378} & 0.808 & 0.404 & 0.188 & 0.444 \\
        GPT-3.5 Turbo, Global & 1.311 & 3.399 & 17.740 & 6.547 & 2.959 & 0.914 & 5.478 \\
        GPT-3.5 Turbo, Local & \textbf{0.207} & \textbf{0.012} & \textbf{0.378} & \textbf{0.383} & 0.316 & \textbf{0.023} & \textbf{0.220} \\
        GPT-3.5 Turbo, Sequential & 0.394 & 0.363 & \textbf{0.378} & 0.596 & \textbf{0.088} & 0.174 & 0.332 \\
        \hline
        Random graph & 0.625 & 1.000 & \textbf{0.125} & 0.750 & 0.750 & 0.465 & 0.619 \\
        Preferential attachment & 0.625 & 0.750 & \textbf{0.125} & 0.750 & 0.750 & 0.416 & 0.569 \\
        Small world & 0.625 & 0.375 & \textbf{0.125} & 0.750	& 0.500 & 0.617 & 0.499 \\
        GPT-3.5 Turbo, Global & 0.967 & 0.933 & 0.833 & 0.900 & 0.967 & 0.740 & 0.890 \\
        GPT-3.5 Turbo, Local & 0.525 & \textbf{0.250} & \textbf{0.125} & \textbf{0.500} & 0.500 & 0.265 & 0.361 \\
        GPT-3.5 Turbo, Sequential & \textbf{0.375} & 0.325 & \textbf{0.125} & 0.625 & \textbf{0.342} & \textbf{0.190} & \textbf{0.330}
    \end{tabular}
    \caption{Quantitative results on structural characteristics. The top six rows indicate difference in means, normalized by the real networks' standard deviation (Eq.~\ref{eqn:norm-mean-diff}). $^\dagger$ indicates that the model parameters were fitted on the real-world mean value for this characteristic, so the mean difference should be ignored, and it is left out of the average. The bottom six rows represent the two-sample Kolmogorov–Smirnov statistic, which measures the distance between two empirical distributions (Eq.~\ref{eqn:ks}). For both measures, lower is better.}
    \label{tab:metrics}
\end{table*}
\begin{table*}
    \centering
    \begin{tabular}{c|c|c}
         \textbf{Study / LLM Method} & \textbf{Demographic} & \textbf{Cross-group ratio (Eq. \ref{eqn:cross-group-ratio})} $\downarrow$ \\
         \hline \citet{thelwall2009homophily}, MySpace & Gender & 1.04 \\
         & Race/ethnicity & 0.44 \\
         & Religion & 0.69 \\
         \hline 
         \citet{laniado2016gender}, Tuenti, friendship & Gender & 0.96 \\
         \citet{laniado2016gender}, Tuenti, interaction & Gender & 0.88 \\
         \hline 
         \citet{smith2014gss}, GSS in 1985 & Gender & 0.81 \\
         & Age & 0.59 \\
         & Race/ethnicity & 0.17 \\
         & Religion & 0.45 \\
         \citet{smith2014gss}, GSS in 2004 & Gender & 0.88 \\
         & Age & 0.60 \\
         & Race/ethnicity & 0.25 \\
         & Religion & 0.44 \\
         \hline 
         \hline 
         Local & Gender & 0.774 (0.015) \\
         & Age & 0.894 (0.010) \\
         & Race/ethnicity & 0.796 (0.013) \\
         & Religion & 0.755 (0.013) \\
         \hline 
         Sequential & Gender & 0.779 (0.017) \\
         & Age & 0.900 (0.013) \\
         & Race/ethnicity & 0.804 (0.018) \\
         & Religion & 0.810 (0.016) \\
         \hline 
    \end{tabular}
    \caption{\rr{Comparing levels of homophily in real-world vs. LLM-generated networks, for the demographic variables besides political affiliation. For Local and Sequential, we report the mean and standard error (in parentheses) over each method's 30 generated networks.}}
    \label{tab:homophily-other-demos}
\end{table*}

In this section, all experiments are run with GPT-3.5 Turbo, unless otherwise specified.

\paragraph{Comparing to classical network models.}
In the main text, we showed that the LLM can match many structural characteristics of real social networks, including density, clustering, connectivity, and degree distribution (Figures~\ref{fig:metrics} and \ref{fig:degree}).
However, how does this compare to existing models for network generation?
Here, we consider three classical network models: (1) Erdős–Rényi random graph models \cite{erdds1959random}, (2) Barabási-Albert preferential attachment models \cite{barabasi1999emergence}, and (3) Watts-Strogatz small-world models \cite{watts1998collective}.
\rr{We choose these models to compare to since they only have 1-2 parameters, and we are interested in LLM's capabilities to generate networks without additional training and with minimal need to defined parameters.}
In contrast, we do not compare to stochastic block models \cite{holland1983stochastic}, \rr{which require edge probabilities between all pairs of blocks and block assignments}, or machine learning models for graph generation, since they have many parameters and require a substantial set of observed graphs for training to fit those parameters \cite{you2018graphrnn,simonovsky2018graphvae,mlgraphgeneration}.

To quantify how well a network metric is matched, we extract the metric from each real social network, using the same eight social networks as in our main experiments (Sections~\ref{sec:structure} and \ref{sec:app-structure}), and from each generated network, with 30 generated networks per model.
To compare the two distributions of the metric, we report both the difference in the means, normalized by the real networks' standard deviation (Eq.~\ref{eqn:norm-mean-diff}), and the two-sample Kolmogorov–Smirnov statistic (Eq.~\ref{eqn:ks}), which measures the distance between two empirical distributions.
As we show in Table~\ref{tab:metrics}, even in the best case---when we allow the models to choose parameters based on the real social networks---these models cannot match all of the real network metrics as well as our Local or Sequential methods can. 
Thus, being able to match the structural characteristics of real social networks is non-trivial, adding significance to our finding that LLMs can match many structural characteristics at once.

\paragraph{Real-world homophily for other demographics.}
In the main text, we showed that levels of political homophily predicted by the LLM are unrealistically high, compared to reported levels of homophily in real-world social networks (Table~\ref{tab:political}).
What about real-world levels of homophily for the other demographic variables?
\rr{Data from \citet{thelwall2009homophily}, who studies connections on MySpace, suggests \rr{cross-group ratios of 0.44 for race/ethnicity, 0.69 for religion, and 1.04 for gender.
Data from \citet{laniado2016gender}, who study connections on Tuenti (a Spanish social network platform), suggests cross-gender ratios of 0.96 when measuring friendship networks and 0.88 when measuring interaction networks.
\citet{smith2014gss}, using data from the US General Social Survey (GSS), finds cross-group ratios of 
0.17 for race/ethnicity, 0.45 for religion, 0.59 for age, and 0.81 for gender in 1985 and ratios of 0.25 for race/ethnicity, 0.44 for religion, 0.60 for age, and 0.88 for gender in 2004}.}

\rr{These numbers, which we summarize in Table \ref{tab:homophily-other-demos}, reveal how much variation there is in levels of homophily across studies and over time, making it difficult to evaluate whether the LLM's levels of homophily are realistic or not.
However, some trends emerge: gender homophily tends to be weaker, race/ethnicity homophily tends to be stronger, and religion and age are somewhere in between. Quotes from summary papers on homophily support this ordering: }
\citet{mcpherson2001birds} say, ``By the time that they are adults, people have friendship and confidant networks that are relatively sex-integrated (at least when compared to other dimensions like race, age, and education),'' and \citet{thelwall2009homophily} says, ``For U.S. friendship in the last century, the key factors, in decreasing order, seem to be race and ethnicity, age, religion, educational level, occupation, and gender.''

\rr{In contrast, in our LLM experiments, the Sequential method predicts similar levels of homophily for gender, race/ethnicity, and religion---all around 0.8---and less homophily for age (0.9) (Table \ref{tab:homophily-other-demos}). 
The Local method predicts slightly more homophily for all variables, but a similar ranking: gender, race/ethnicity, and religion have cross-group ratios around 0.75 and age still has less homophily (0.89). 
While these levels of homophily seem realistic for gender, it seems that the LLM seriously underestimates homophily for race/ethnicity and religion (and likely for age as well, although this is inconclusive since we only have age homophily from one study and age homophily is less comparable to the others since it is not a categorical variable).
It is possible that LLMs, due to their guardrails and instruction fine-tuning, have been guided away from using sensitive attributes like race or religion to influence their generation; however, future work is needed to thoroughly investigate evidence of underestimation and probe for possible reasons.
}

\paragraph{Age homophily.}
\begin{figure}[t]
    \centering
    \includegraphics[width=\linewidth]{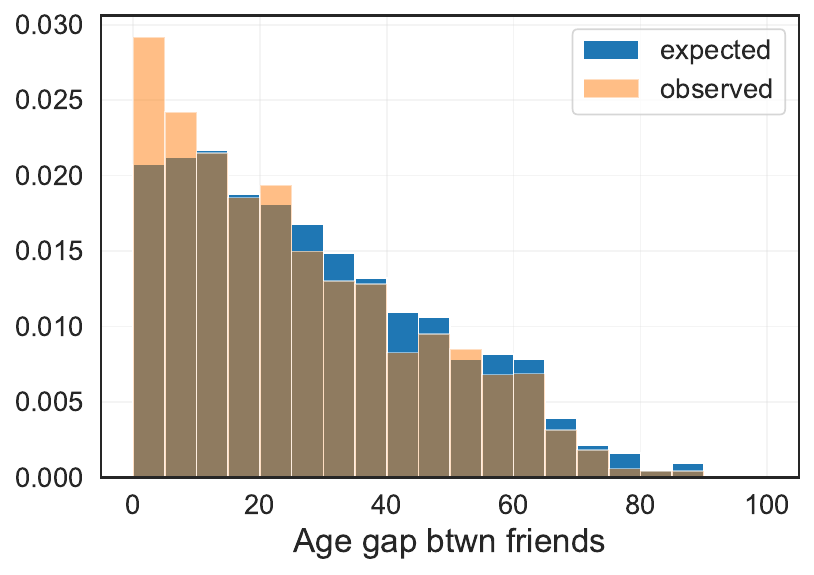}
    \caption{Observed (left) vs. expected (right) distribution of age gaps in the generated social networks, under the Sequential method.}
    \label{fig:age-homophily}
\end{figure}
\rr{Age is our only non-categorical demographic variable, which slightly complicates our definition of homophily, which assumes well-defined groups. We are able to measure an approximately analogous cross-group ratio (Eq. \ref{eqn:cross-group-ratio}) for age by comparing the observed-to-expected average age gaps between connected nodes, as described in the main text. 
As a more detailed measure, we can also plot the distribution of observed age gaps vs. expected age gaps, which we show in Figure~\ref{fig:age-homophily}.
The observed distribution consists of the age gaps for all observed edges pooled over the 30 networks generated by the Sequential method. The expected distribution consists of the age gaps between all possible pairs of nodes among the 50 personas.}
Compared to the expected distribution, the observed distribution is clearly shifted to the left (i.e., smaller gaps), with more edges than expected with age gaps of 0-10. 

\paragraph{Ablations of demographic variables.}
\begin{figure*}
    \centering
    \includegraphics[width=\linewidth]{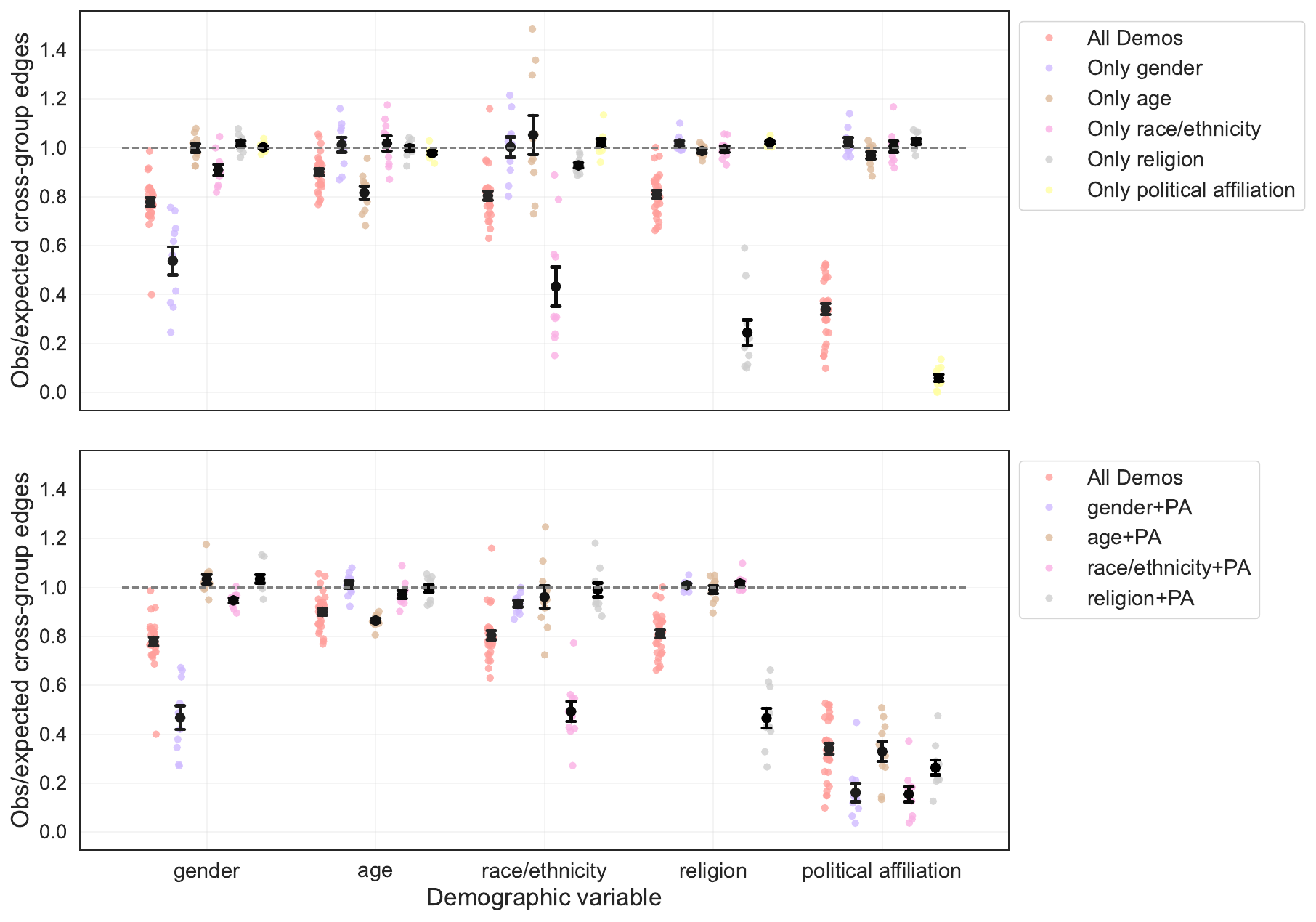}
    \caption{
    Evaluating homophily per demographic variable, when providing subsets of the demographics.
    \textbf{Top:} providing one variable at a time. \textbf{Bottom:} providing two variables at a time, political affiliation (PA) and one of the other four. Ratios \rr{below} 1 (marked by the grey line) indicate homophily and \rr{lower} ratios indicate more homophily.
    In both plots, we visualize mean and standard error (in black) and individual data points corresponding to each network.
    All model results shown here use the Sequential method and GPT-3.5-Turbo.
    }
    \label{fig:ablations}
\end{figure*}
\rr{Here, we explore how the LLM's generated networks change when we provide subsets of the original set of five demographic variables (gender, age, race/ethnicity, religion, and political affiliation). First, we try providing one variable at a time. When only one variable is provided, we find that homophily for that variable gets stronger, compared to when all five variables are provided (Figure \ref{fig:ablations}, top). This is not entirely surprising: when the LLM is only provided this variable, then it forms the social network based entirely on that variable. For example, if the LLM is only given gender, the LLM is unlikely to connect a woman and a man since they do not have any listed demographics in common, but if more demographics were provided, they might now have a demographic in common. However, this result is not guaranteed: the LLM could have learned the joint distribution of these variables during pretraining, so theoretically it could have added the missing demographics to each persona by sampling from that joint distribution. Given that the LLM does not seem to do this on its own, it would be interesting in future work to explore this capability: adding greater variance in the generated networks by augmenting each persona with additional traits, generated by the LLM. However, one would have to be careful about exacerbating biases with this technique, as we saw with LLM-generated interests.}

\rr{Second, we try providing two variables at a time: political affiliation and one of the four other variables. For each variable, its level of homophily when in a pair tends to be between its level of homophily when alone versus with all five variables (Figure \ref{fig:ablations}, bottom). We also find in both of these experiments that political affiliation continues to be the dominant factor. In the single-variable experiments, political homophily is the strongest when political affiliation is provided, compared to homophily for any other variable when it is the only variable provided. In the two-variable experiments where political affiliation and one other variable is provided, political homophily is always stronger than the other variable's homophily.}

\paragraph{Degree per demographic group.}
\begin{figure*}
    \centering
    \includegraphics[width=\linewidth]{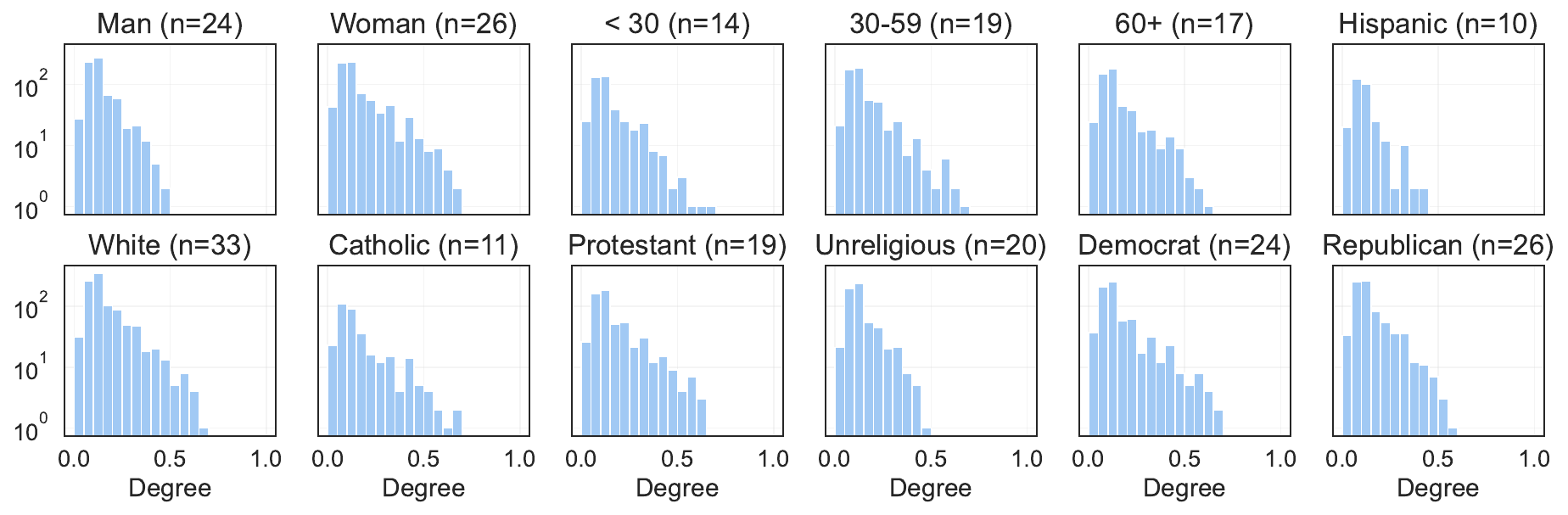}
    \caption{\rr{Degree distribution per demographic group, pooled over the 30 networks generated by the Sequential method.}}
    \label{fig:degree-per-group}
\end{figure*}
\begin{table}[]
    \small
    \centering
    \begin{tabular}{r|c|c|c|c}
         & $n$ & \textbf{Mean degree} & \textbf{Gini coef.} & \textbf{GS Index} \\
         \hline
         Overall & 50 & 0.149 (0.002) & 0.347 (0.005) & -- \\
         \hline 
         Man & 24 & 0.131 (0.003) & 0.281 (0.009) & 0.368 (0.011) \\
         Woman & 26 & 0.166 (0.004) & 0.370 (0.008) & 0.299 (0.009) \\
         \hline 
         Under 30 & 14 & 0.147 (0.005) & 0.321 (0.010) & 0.524 (0.009) \\
         30-59 & 19 & 0.150 (0.004) & 0.336 (0.009) & 0.543 (0.007) \\
         60+ & 17 & 0.150 (0.005) & 0.328 (0.009) & 0.541 (0.006) \\
         \hline 
         Hispanic & 10 & 0.114 (0.004) & 0.259 (0.013) & 0.426 (0.017) \\
         White & 33 & 0.160 (0.003) & 0.335 (0.008) & 0.330 (0.008) \\
         \hline 
         Catholic & 11 & 0.157 (0.006) & 0.354 (0.013) & 0.511 (0.009) \\
         Protestant & 19 & 0.163 (0.004) & 0.352 (0.010) & 0.475 (0.011) \\
         Unreligious & 20 & 0.132 (0.004) & 0.276 (0.008) & 0.487 (0.009) \\
         \hline 
         Democrat & 24 & 0.157 (0.004) & 0.355 (0.009) & 0.189 (0.011) \\
         Republican & 26 & 0.143 (0.005) & 0.313 (0.007) & 0.210 (0.014) \\
    \end{tabular}
    \caption{\rr{Degree distribution and Gini-Simpson (GS) diversity index per demographic group, for all groups with at least 10 personas in our set of 50 personas. For each statistic, we compute its value per network, and report the mean value and standard error (in parentheses) over the 30 networks generated by the Sequential method.}}
    \label{tab:degree-per-group}
\end{table}
\rr{In the main text, we discussed and visualized degree distributions over all nodes in the network (Figure \ref{fig:degree}), but degree could also differ across demographic groups. In Figure \ref{fig:degree-per-group}, we visualize each group's degree distribution, for all groups with at least 10 personas in our set of 50 personas. As in Figure \ref{fig:degree}, the distribution consists of degrees for all nodes in the group, pooled over the 30 networks generated by the Sequential method.
We see minor differences between groups: for example, in the generated networks, women have slightly higher degree on average than men (0.166 vs. 0.131, Table \ref{tab:degree-per-group}).
This could be explained by two factors: first, there are slightly more women ($n$=26) than men ($n$=24) in our set of 50 personas, and second, women show higher levels of same-group preference, as discussed in the main text (same-group ratios of 1.43 vs 0.99, Figure \ref{fig:pairwise-ratios}). 
We also find that White personas have slightly higher average degree than Hispanic personas (0.160 vs 0.114), which could also be explained by there being more White personas ($n$=33) vs. Hispanic personas ($n$=10), and higher same-group ratios among White personas compared to Hispanic personas (1.20 vs. 1.02).}

\rr{Beyond using the mean to describe the degree distribution, we can also use the Gini coefficient as a measure of the inequality in degree.
For a set of values $x_1, x_2, \cdots, x_n$, the Gini coefficient is defined as
\begin{align}
    G = \frac{\sum_{i=1}^n \sum_{j=1}^n |x_i - x_j|}{2n^2 \bar{x}},
\end{align}
where $\bar{x}$ is the mean of the values.
First, when we compute the Gini coeffient over all nodes, we find a higher coefficient (more inequality) when using the Sequential method ($G=0.347$), compared to the Local method ($G=0.261$) or Global method ($G=0.255$).
This aligns with our earlier results showing that Sequential can better capture the long-tailed degree distribution of real social networks. Per group, we find similar results to what we found for degree distribution: higher average degree tends to correspond to higher Gini coefficient, since the higher average tends to be driven by longer tails, which results in greater inequality. As described in Table \ref{tab:degree-per-group}, women compared to men, White compared to Hispanic, Protestant and Catholic compared to Unreligious, and Democrat compared to Republican all have slightly higher average degrees and Gini coefficients.}

\paragraph{Diversity per demographic group.}
\rr{We can also measure rates of homophily and diversity per group. In Figure \ref{fig:pairwise-ratios}, we visualized observed-to-expected ratios to capture same-group preferences (the diagonal) and cross-group preferences (the off-diagonal). This analysis revealed that not all cross-group relations are equally unlikely; for example, adjacent age groups were likelier to have cross-group relations. It also revealed that not all same-group preferences were equally strong; for example, we saw stronger preferences within women compared to men or within Catholic compared to Unreligious.}

\rr{Related to homophily, we can also measure the \textit{diversity} in a persona's 1-hop neighbors (i.e., their friends).
We use the Gini-Simpson (GS) index to measure diversity, which is defined in terms of a set of groups, $\mathcal{G} = \{g_1, g_2, \cdots, g_k\}$, and the proportion $p_g$ of values that belong to each group:
\begin{align}
    GS = 1 - \sum_{g \in \mathcal{G}} p_g.
\end{align}
For a given node $i$, we define its 1-hop diversity with respect to a demographic variable by computing the GS index over their friends' group identities for that variable.
In Table \ref{tab:degree-per-group}, we report the average 1-hop diversity for members in a group, with respect to that group's demographic variable (e.g., the average 1-hop gender diversity for men).
As expected, groups with higher diversity correspond to those with lower same-group ratios in Figure \ref{fig:pairwise-ratios}: men compared to women, Hispanic compared to White, Catholic compared to Protestant and Unreligious, and Republican compared to Democrat.
The three age groups are similar to each other for both measures.}

\paragraph{Individuals and intersections of group identities.}
\rr{We found that LLMs overemphasize political homophily overall and that each political group, Democrats and Republicans, shows strong same-group preferences (Figure \ref{fig:pairwise-ratios}) and low diversity in 1-hop neighbors (Table \ref{tab:degree-per-group}). However, how does political preference and diversity vary over individuals, as their other group identities intersect with their political affiliation? 
To analyze this, we use individual-level metrics: the GS index to measure individual-level diversity in 1-hop neighbors and an individual's``Democrat lean'', i.e., the proportion of the individual's 1-hop neighbors who are Democrats, as an easily interpretable measure.}

\rr{We find large within-group variability over individuals: for example, the persona with the lowest 1-hop political diversity (measured by GS index) is a woman, age 75, White, Unreligious, and Democrat (mean=0.111, SE=0.032), while the persona with the highest 1-hop political diversity is a woman, age 38, White, Protestant, and Democrat (mean=0.353, SE=0.028). Relatedly, we find that the lowest-diversity persona has a very high Democrat lean (mean=0.908, SE=0.030), while the highest-diversity persona’s lean is less extreme (mean=0.722, SE=0.028). 
Since both individuals are women, White, and Democrat, we can attribute their differences to their age and religion. From pure numbers, we would not expect religion to play a major role in their differences, since the share of Democrats vs. Republicans is about the same for personas that identify as Unreligious (10 vs. 10) and as Protestant (10 vs. 9). However, it is possible that the LLM expects that an Unreligious Democrat’s preference for befriending Democrats is stronger than a Protestant Democrat’s preference. 
Age could also partially explain their differences: among personas within 10 years of 75 (65-85), there are 6 Democrats and 3 Republicans, but among personas within 10 years of 38 (28-48), there are 8 Republicans and 5 Democrats, which could help to explain why the former persona has a stronger Democrat lean and less political diversity in friends.}

\rr{These results reveal that, even though we see strong political homophily and low political diversity overall, the LLM generates some variability across individuals, due to the intersection of their political affiliation with other demographic identities, such as religion or age.}

\paragraph{Incorporating interests.}
\begin{figure*}
    \centering
    \includegraphics[width=0.8\linewidth]{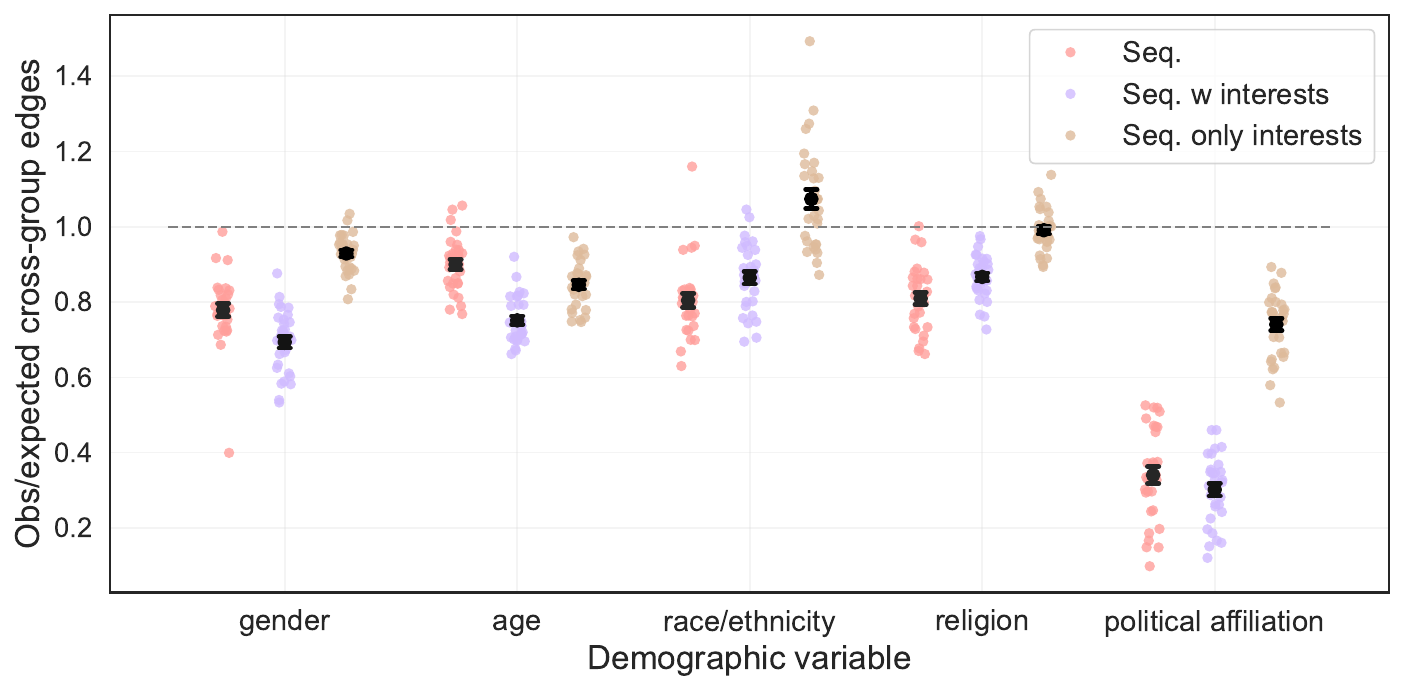}
    \caption{Visualizing demographic homophily under three methods: ``Seq.'' (only demographic information), ``Seq. w interests'' (demographic information and interests), and ``Seq. only interests'' (only interests). The LLM places the largest emphasis on political homophily in all cases. Ratios \rr{below} 1 (marked by the grey line) indicate homophily, with \rr{lower} ratios indicating more homophily.
    We visualize mean and standard error (in black) and individual data points corresponding to each network.
    All model results shown here use the Sequential method and GPT-3.5-Turbo.}
    \label{fig:only-interests}
\end{figure*}
\begin{table*}[t]
    \small
    \centering
    \begin{tabular}{p{2.5cm}|p{12cm}}
       \textbf{Demographic group} & \textbf{Top interests} \\
       \hline 
       Man & social justice (29.1\%), church activities (23.8\%), conservative politics (23.4\%), sports (17.1\%), community service (16.7\%), technology (15.7\%), golf (14.0\%), history (13.6\%), video games (12.6\%), progressive policies (11.2\%) \\
       Woman & social justice (35.3\%), gardening (31.5\%), community service (27.7\%), reading (21.6\%), church activities (20.8\%), conservative politics (16.2\%), volunteering (12.8\%), family gatherings (11.6\%), travel (10.8\%), cooking (9.2\%) \\
       \hline 
       White & gardening (27.4\%), conservative politics (25.8\%), church activities (24.0\%), social justice (23.0\%), community service (15.9\%), reading (15.6\%), outdoor activities (12.0\%), history (11.0\%), volunteering (11.0\%), golf (10.7\%) \\
       Hispanic & social justice (46.0\%), community service (35.8\%), family gatherings (27.3\%), church activities (18.2\%), cultural heritage (16.0\%), music (13.4\%), progressive policies (12.3\%), conservative politics (12.3\%), family (10.7\%), soccer (10.7\%) \\
       Black & social justice (58.6\%), community service (33.1\%), church activities (21.1\%), gospel music (17.3\%), progressive policies (15.8\%), community activism (14.3\%), music (14.3\%), reading (13.5\%), jazz music (11.3\%), sports (11.3\%) \\
       Asian & social justice (27.9\%), travel (24.6\%), technology (23.0\%), church activities (16.4\%), community service (14.8\%), conservative politics (14.8\%), yoga (13.1\%), fitness (11.5\%), entrepreneurship (11.5\%), gardening (11.5\%) \\
       \hline 
       Protestant & church activities (36.1\%), social justice (33.2\%), community service (25.2\%), gardening (22.0\%), conservative politics (21.4\%), reading (13.9\%), sports (10.9\%), volunteering (9.5\%), history (9.3\%), outdoor activities (8.9\%) \\
       Catholic & community service (37.7\%), social justice (35.2\%), family gatherings (25.5\%), church activities (24.3\%), conservative politics (21.5\%), gardening (16.2\%), reading (10.1\%), sports (9.3\%), volunteering (8.9\%), progressive policies (8.5\%) \\
       Unreligious & social justice (28.3\%), technology (22.8\%), gardening (21.0\%), travel (18.8\%), fitness (16.3\%), conservative politics (15.9\%), reading (13.0\%), outdoor activities (13.0\%), progressive policies (12.7\%), entrepreneurship (9.4\%) \\
       \hline 
       Republican & conservative politics (41.6\%), church activities (32.1\%), gardening (23.2\%), outdoor activities (15.0\%), golf (14.3\%), community service (13.9\%), sports (13.5\%), family gatherings (12.7\%), history (12.2\%), hunting (11.4\%) \\
       Democrat & social justice (62.5\%), community service (29.3\%), progressive policies (18.6\%), gardening (16.4\%), reading (15.6\%), church activities (13.2\%), volunteering (12.4\%), music (11.0\%), travel (10.4\%), technology (10.0\%)
    \end{tabular}
    \caption{Top 10 interests per demographic group.}
    \label{tab:top-interests-per-demo}
\end{table*}
In Figure~\ref{fig:only-interests}, we visualize demographic homophily under three versions of the Sequential method: when only demographic information is provided, when demographic information and interests are provided, and when only interests are provided.
In all cases, the LLM places the largest emphasis on political homophily.

In Table~\ref{tab:top-interests-per-demo}, we report the top interests per demographic group.
The LLM generates interests as a comma-separated list (see examples in Table~\ref{tab:interests}), which makes it straightforward to separate the list into individual interests and compute, for each group and interest, what percentage of personas in that group have that interest. 
For this experiment, we use 1,000 personas, instead to our usual set of 50 personas, so that we can estimate percentages over larger populations.
We report percentages for all groups with at least 30 personas (see the counts per group in Table~\ref{tab:personas}).
We find that ``social justice'' is a common interest across groups since it dominates Democrats' interests (62.5\%) and Democrats account for a large portion of every other group besides Republicans.
We can also see the result of correlations between demographic identities: for example, men, White, and religious populations are likelier to be Republicans, resulting in higher percentages for ``conservative politics''.

\paragraph{Sensitivity analyses.}
\begin{figure*}
    \centering
    \includegraphics[width=0.8\linewidth]{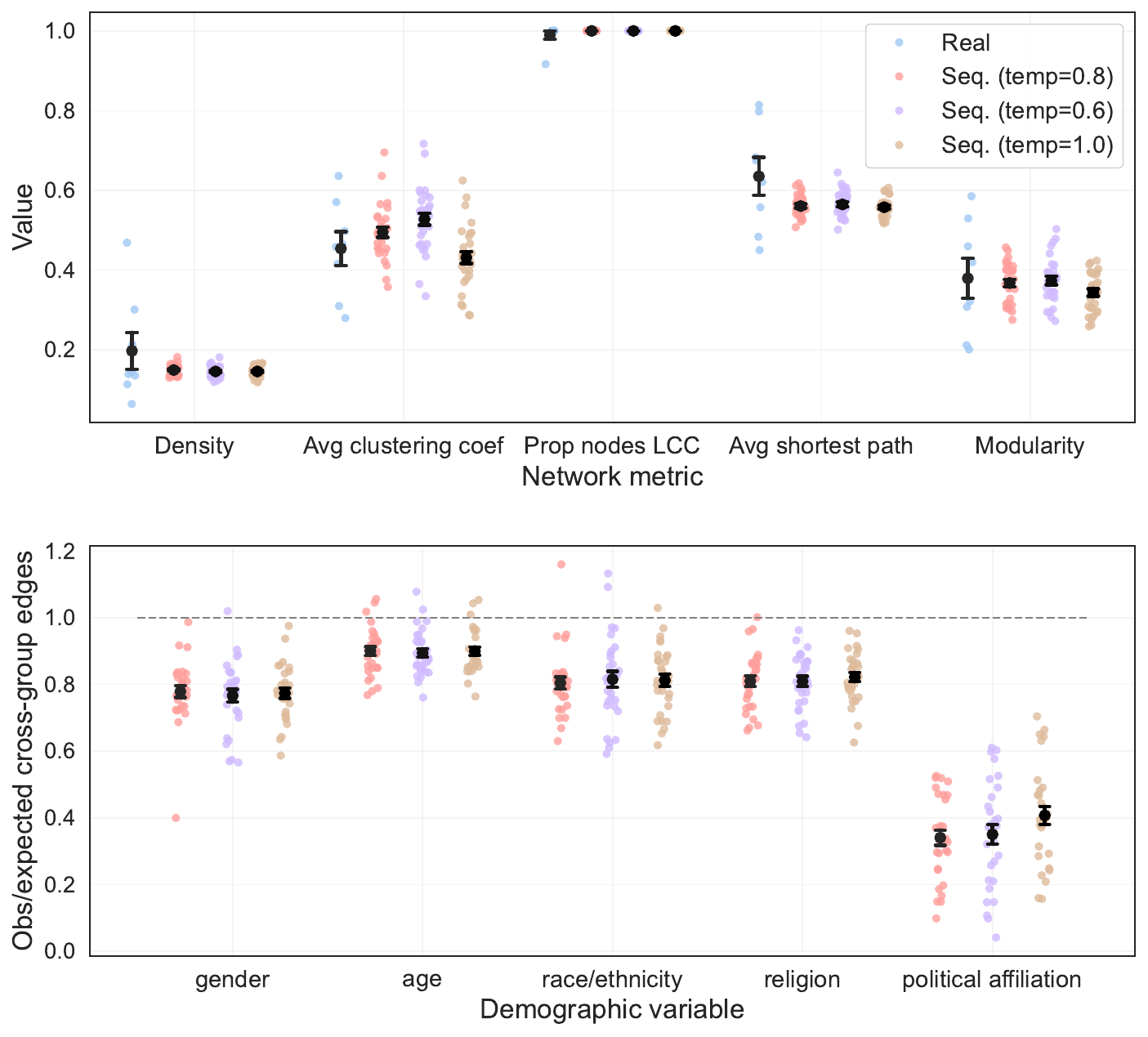}
    \caption{
    In our experiments, we use a default temperature of 0.8. Our main results do not change significantly if we use a temperature of 0.6 or 1.0 instead.
    \textbf{Top:} structural network metrics. \textbf{Bottom:} homophily, where ratios \rr{below} 1 (marked by the grey line) indicate homophily and \rr{lower} ratios indicate more homophily.
    In both plots, we visualize mean and standard error (in black) and individual data points corresponding to each network. 
    All model results shown here use the Sequential method and GPT-3.5-Turbo.
    }
    \label{fig:comparing-temp}
\end{figure*}

\begin{figure*}
    \centering
    \includegraphics[width=\linewidth]{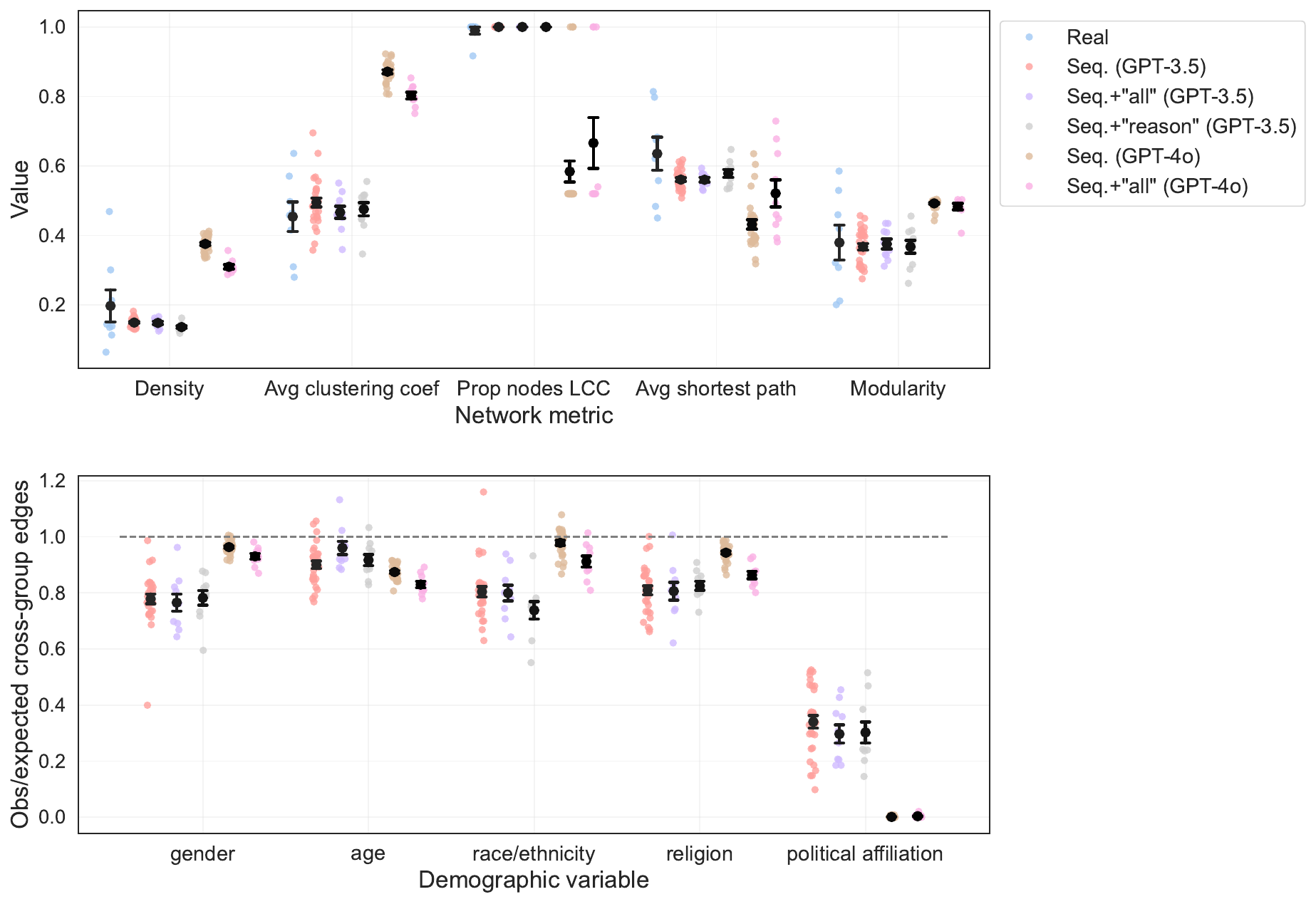}
    \caption{
    Our results do not change siginificantly with minor changes to the prompt. We try adding ``Pay attention to all demographics'' (+``all'') and prompting the LLM to give a short reason for each friend that it selects (+``reason'').
    \textbf{Top:} structural network metrics. \textbf{Bottom:} homophily, where ratios \rr{below} 1 (marked by the grey line) indicate homophily and \rr{lower} ratios indicate more homophily.
    In both plots, we visualize mean and standard error (in black) and individual data points corresponding to each network.
    All model results shown here use the Sequential method.
    }
    \label{fig:prompt-change}
\end{figure*}

In our experiments, we use a default temperature of 0.8. 
In Figure~\ref{fig:comparing-temp}, we show that our main results do not significantly change if we use a temperature of 0.6 or 1.0 instead.
Since the LLM overemphasizes political homophily, we also try adding ``Pay attention to all demographics'' to the system prompt.
We include GPT-4o in this study as well due to its extreme levels of political homophily (Figure~\ref{fig:gpt4o}).
In Figure~\ref{fig:prompt-change}, we show that adding this prompt does not significantly change results for GPT-3.5 Turbo or GPT-4o.
Finally, as mentioned before, we also find that results do not significantly change if we prompt the LLM to generate a short reason for each friend that it selects.
Thus, our results are robust to these perturbations in temperature and prompt.

\setcounter{figure}{0}
\setcounter{table}{0}
\section{Methodological Details}
\label{sec:app-methods}

\subsection{Persona construction}
\label{sec:persona-details}
\begin{table*}
    \centering
    \begin{tabular}{c|c|c|c}
       \textbf{Demographic} & \textbf{Group} & \textbf{Count in 50} & \textbf{Count in 1,000} \\
       \hline 
       Gender & Woman & 26 & 499 \\
        & Man & 24 & 492 \\
        & Nonbinary & 0 & 9 \\
        \hline 
        Race/ethnicity & White & 33 & 609 \\
        & Hispanic & 10 & 187 \\
        & Black & 4 & 133 \\
        & Asian & 2 & 61 \\
        & American Indian/Alaska Native & 1 & 9 \\
        & Native Hawaiian/Pacific Islander & 0 & 1 \\
        \hline 
        Religion & Protestant & 19 & 440 \\
        & Unreligious & 20 & 276 \\
        & Catholic & 11 & 247 \\
        & Jewish & 0 & 12 \\
        & Hindu & 0 & 9 \\
        & Buddhist & 0 & 8 \\
        & Muslim & 0 & 6 \\
        & Other Christian & 0 & 2 \\
        \hline 
        Political affiliation & Democrat & 24 & 501 \\ 
        & Republican & 26 & 474 \\
        & Independent & 0 & 25
    \end{tabular}
    \caption{Marginal distributions of personas' demographics. Count in 50 indicates the number of personas belonging to each group in the sample of 50 personas we used for most experiments. Count in 1,000 indicates their number in the sample of 1,000 personas we used for evaluating top interests per demographic group (Table~\ref{tab:top-interests-per-demo})}
    \label{tab:personas}
\end{table*}
\begin{figure}[t]
    \centering
    \includegraphics[width=\linewidth]{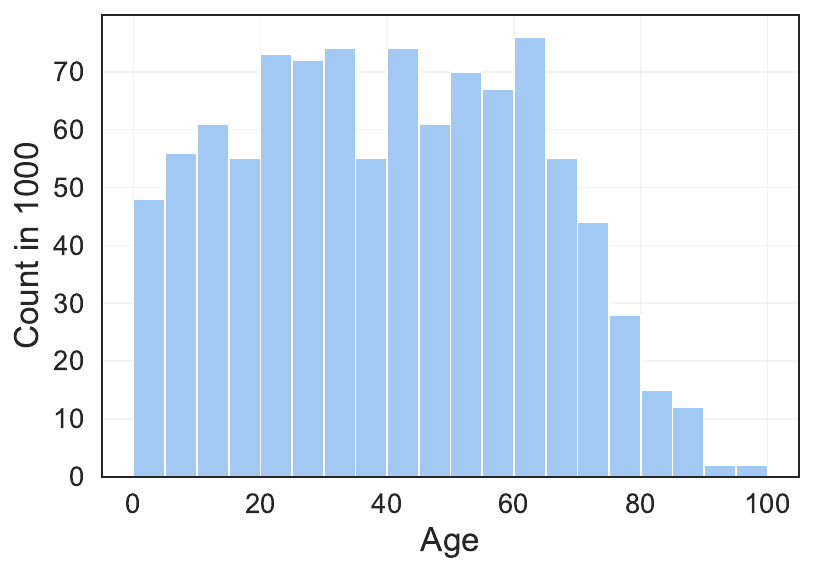}
    \caption{Age distribution in sample of 1,000 personas.}
    \label{fig:age-dist}
\end{figure}
As described in the main text, we include gender, age, race/ethnicity, religion, and political affiliation.
In Table~\ref{tab:personas}, we list the number of personas in each demographic group for the sample of 50 personas we used in most of our experiments, as well as the sample of 1,000 personas we used for evaluating top interests per demographic (Table~\ref{tab:top-interests-per-demo}).
In Figure~\ref{fig:age-dist}, we visualize the distribution of ages in the sample of 1,000 personas.
Below, we explain how we sampled demographic variables per persona.

\paragraph{Gender, race/ethnicity, and age.}
First, we use data from the US Census \cite{census2023joint}, who provide monthly population estimates for sex, race/ethnicity, and age (individual years, from 0 to 100 years old). 
Specifically, we downloaded \texttt{nc-est2023-alldata-r-file07.csv} from US Census datasets archived online\footnote{\url{https://www2.census.gov/programs-surveys/popest/datasets/2020-2023/national/asrh/}} and used the data for June 2023.
We use these estimates to calculate joint distributions of gender, race/ethnicity, and age.
Additionally, using data from Pew Research Center \cite{pew2022trans}, we sample from the age-dependent distribution of those who identify as non-binary.

\paragraph{Religion.}
We sample religion conditioned on the persona's race/ethnicity.
Statista \cite{statista2016racereligion} provides distributions of religious identity for adults in the US in 2016, for most race/ethnicities.
Additionally, using data from 2020 PRRI Census of American Religion \cite{prri2020religious}, we acquire the distribution for Native Americans.

\paragraph{Political affiliation.}
Finally, we sample political affiliation conditioned on the persona's race/ethnicity and gender. 
We primarily use data from Pew Research Center \cite{pew2024political}, using the 2023 numbers from their figure, ``Partisan identification by gender among racial and ethnic groups,'' which cover most race/ethnicities.
Additionally, we use data from Brookings \cite{brookings2022native}, who report Native Americans' distribution of political support in 2022.

\paragraph{Interests.}
\begin{figure*}
    \centering
    \begin{tikzpicture}
        \node[draw,text width=15cm] at (0,0) {\textbf{User:} In 8-12 words, describe the interests of someone with the following demographics:\\
        race/ethnicity: White\\
        age: 72\\
        gender: Man\\
        political affiliation: Republican\\
        religion: Catholic\\
        Answer by providing ONLY their interests. Do not include filler like ``She enjoys'' or ``He has a keen interest in''.};
    \end{tikzpicture}
    \caption{Prompt to generate interests for persona.}
    \label{fig:interest-prompt}
\end{figure*}
In Figure~\ref{fig:interest-prompt}, we provide the prompt that we use to generate interests.
We randomize the order of demographics provided, since we find that the LLM seems to pay special attention to the first listed demographic when generating interests.
We use GPT-4o to generate interests, since we find that it follows the required format a little better.

\subsection{Network generation}
\label{sec:app-network}
\begin{figure*}
    \centering
    \begin{tikzpicture}
        \node[draw,text width=15cm] at (0,0) {\textbf{System:} Your task is to create a realistic social network. You will be provided a list of people in the network, where each person is described as ``ID. Gender, Age, Race/ethnicity, Religion, Political affiliation''. Provide a list of friendship pairs in the format ID, ID with each pair separated by a newline. Do not include any other text in your response. Do not include any people who are not listed below.};
        \node[draw,text width=15cm] at (0,-2.8) {\textbf{User:} 28. Man, age 48, Hispanic, Protestant, Democrat\\
        11. Man, age 31, White, Protestant, Democrat\\
        10. Man, age 58, Hispanic, Catholic, Democrat\\
        41. Woman, age 41, White, Catholic, Republican\\
        ...};
    \end{tikzpicture}
    \caption{Prompt for Global method.}
    \label{fig:global-prompt}
\end{figure*}
\begin{figure*}
    \centering
    \begin{tikzpicture}
        \node[draw,text width=15cm] at (0,0) {\textbf{System:} You are a Man, age 48, Hispanic, Protestant, Democrat. You are joining a social network.
        You will be provided a list of people in the network, where each person is described as ``ID. Gender, Age, Race/ethnicity, Religion, Political affiliation''.
        Which of these people will you become friends with? Provide a list of *YOUR* friends in the format ID, ID, ID, etc. Do not include any other text in your response. Do not include any people who are not listed below.};
        \node[draw,text width=15cm] at (0,-3) {\textbf{User:} 11. Man, age 31, White, Protestant, Democrat\\
        10. Man, age 58, Hispanic, Catholic, Democrat\\
        41. Woman, age 41, White, Catholic, Republican\\
        2. Woman, age 20, White, Catholic, Republican\\
        ...};
    \end{tikzpicture}
    \caption{Prompt for Local method.}
    \label{fig:local-prompt}
\end{figure*}
\begin{figure*}
    \centering
    \begin{tikzpicture}
        \node[draw,text width=15cm] at (0,0) {\textbf{System:} You are a Man, age 48, Hispanic, Protestant, Democrat. You are joining a social network.
        You will be provided a list of people in the network, where each person is described as ``ID. Gender, Age, Race/ethnicity, Religion, Political affiliation'', followed by their current number of friends.
        Which of these people will you become friends with? Provide a list of *YOUR* friends in the format ID, ID, ID, etc. Do not include any other text in your response. Do not include any people who are not listed below.};
        \node[draw,text width=15cm] at (0,-3) {\textbf{User:} 11. Man, age 31, White, Protestant, Democrat; has 4 friends\\
        10. Man, age 58, Hispanic, Catholic, Democrat; has 2 friends\\
        41. Woman, age 41, White, Catholic, Republican; has 0 friends\\
        2. Woman, age 20, White, Catholic, Republican; has 7 friends\\
        ...};
    \end{tikzpicture}
    \caption{Prompt for Sequential method, when only degree is provided. We also have a version where each persona's current list of friends (in the form of ID) is provided.}
    \label{fig:sequential-prompt}
\end{figure*}

In Figures~\ref{fig:global-prompt}-\ref{fig:sequential-prompt}, we provide the full basic prompts for each of our network generation methods: Global, Local, and Sequential.
These are the prompts that are used when only demographic variables are provided per persona.
When interests are provided, we add ``interests include: ...'' per persona.
In variants of the Sequential prompt, discussed in Appendices~\ref{sec:diff-llms} and \ref{sec:app-results}, we experiment with specifying the number of friends that should be chosen; prompting the LLM to generate a short reason for each selected friend; and adding ``Pay attention to all demographics'' to the prompt.

For the Global method, the entire list of personas is given in one prompt, while in the Local and Sequential methods, the LLM is assigned one persona at the time and all other personas are listed.
We generate 30 networks per method where, for each generated network, we randomize the order that the personas are listed, and, for the Local and Sequential methods, we also randomize the order in which personas are assigned (using different orders for listing and assignment). 
For the Sequential method, we experiment with providing each persona's list of friends versus only their degree.
We find that the model performs better with only degree, while listing friends results in unrealistically high densities. 
Furthermore, fewer tokens are used with only degree, so we use this version of Sequential.

\rr{In Figure \ref{fig:networks}, we visualize examples of networks generated by each of the three methods.
The networks are visualized using \texttt{networkx}, with a spring layout and fixed seed for the visualization.
The visualizations primarily reveal differences in the generated networks across methods, but they also reveal differences between networks generated by the same method.
Since the visualization method is fixed, these differences should be attributed to variation in actual networks, driven by the randomization in persona ordering (for listing and assignment) and randomness in LLM outputs (using a default temperature of 0.8).}

\paragraph{Model costs and comprehension.}
\begin{table*}[]
    \centering
    \begin{tabular}{p{3.2cm}|p{2.8cm}|p{1.8cm}|p{1.8cm}|p{1.8cm}|p{1.8cm}}
        \textbf{Method} & $O$(\textbf{input}) & \textbf{\# input tokens} & \textbf{\# output tokens} & \textbf{Expected \# turns} & \textbf{Actual \# turns} \\
        \hline 
        Global, demos & $O(ND)$ & 607 & 95.267 & 1 & 1.133 \\
        Local, demos & $O(N^2D)$ & 30450 & 196.767 & 50 & 50 \\
        Sequential (friend list), demos & $O(N^2D+NE)$ & 42601.233 & 335.633 & 50 & 50 \\
        Sequential (degree), demos & $O(N^2D)$ & 37735 & 200.300 & 50 & 50.033 \\
        Sequential (degree), interests & $O(N^2I)$ & 37035 & 257.967 & 50 & 50.033 \\
        Sequential (degree), demos + interests & $O(N^2(D+I))$ & 57385 & 190.733 & 50 & 50.067 
    \end{tabular}
    \caption{Comparing model costs and comprehension over different LLM methods for social network generation. For Sequential, we consider two versions: listing each persona's current friend list vs. only their current degree. $N$ is the number of nodes in the network, $E$ is the number of edges, and $D$ and $I$ are the number of tokens needed to describe all demographics and interests, respectively.}
    \label{tab:costs}
\end{table*}
In this work, we focus primarily on realism and bias to evaluate the generated networks.
However, when using LLMs, measures such as token costs and model comprehension are also very important.
We measure token cost as the number of input tokens and output tokens, summed over generating the entire network.
We also conduct a big-O analysis of how the input tokens scale with network size, based on $N$ (the number of nodes in the network), $E$ (the number of edges), $D$ (the number of tokens to describe \rr{a persona's} demographics), and $I$ (the number of tokens to describe \rr{a persona's} interests).
To measure model comprehension, we consider the expected number of turns (prompt and response) to generate the entire network, compared to the actual number of turns.
The actual exceeds the expected if we could not parse the model's response (e.g., it gave an ID that does not exist in the network).
So that we can measure the concepts of cost and model comprehension independently, we only add to the number of input/output tokens on \textit{successful} turns.

We report results in Table~\ref{tab:costs}, with all measures averaged over the 30 networks generated per method.
The Global method requires, by far, the fewest number of input tokens, since we only need to list all personas once.
However, Global has the highest rate of errors, resulting in an actual number of turns that is 13\% higher than the expected number of turns.
The Local method and Sequential method scale similarly, since they both require iterating through all $N$ personas, and for each persona, presenting them with information about the other $N-1$ personas.
Their main difference is what information is provided about each persona: \rr{Local only provides $D$ demographics, while Sequential also provides network information, which can be each persona's friend list (resulting in an additional $E$ per iteration) or only each persona's degree (resulting in $D+1$ per persona)}. 
Both Local and Sequential have good model comprehension, with the models rarely failing and actual turns being only 0-6.7\% above expected turns.
Overall, this table lays out tradeoffs, since the Global method requires far fewer tokens, but as we saw in the main text (Figures~\ref{fig:networks}-\ref{fig:degree}), it produces much less realistic networks, and it is more prone to invalid responses. 

\paragraph{Scalability.} 
\begin{figure*}[t]
    \centering
    \includegraphics[width=0.75\linewidth]{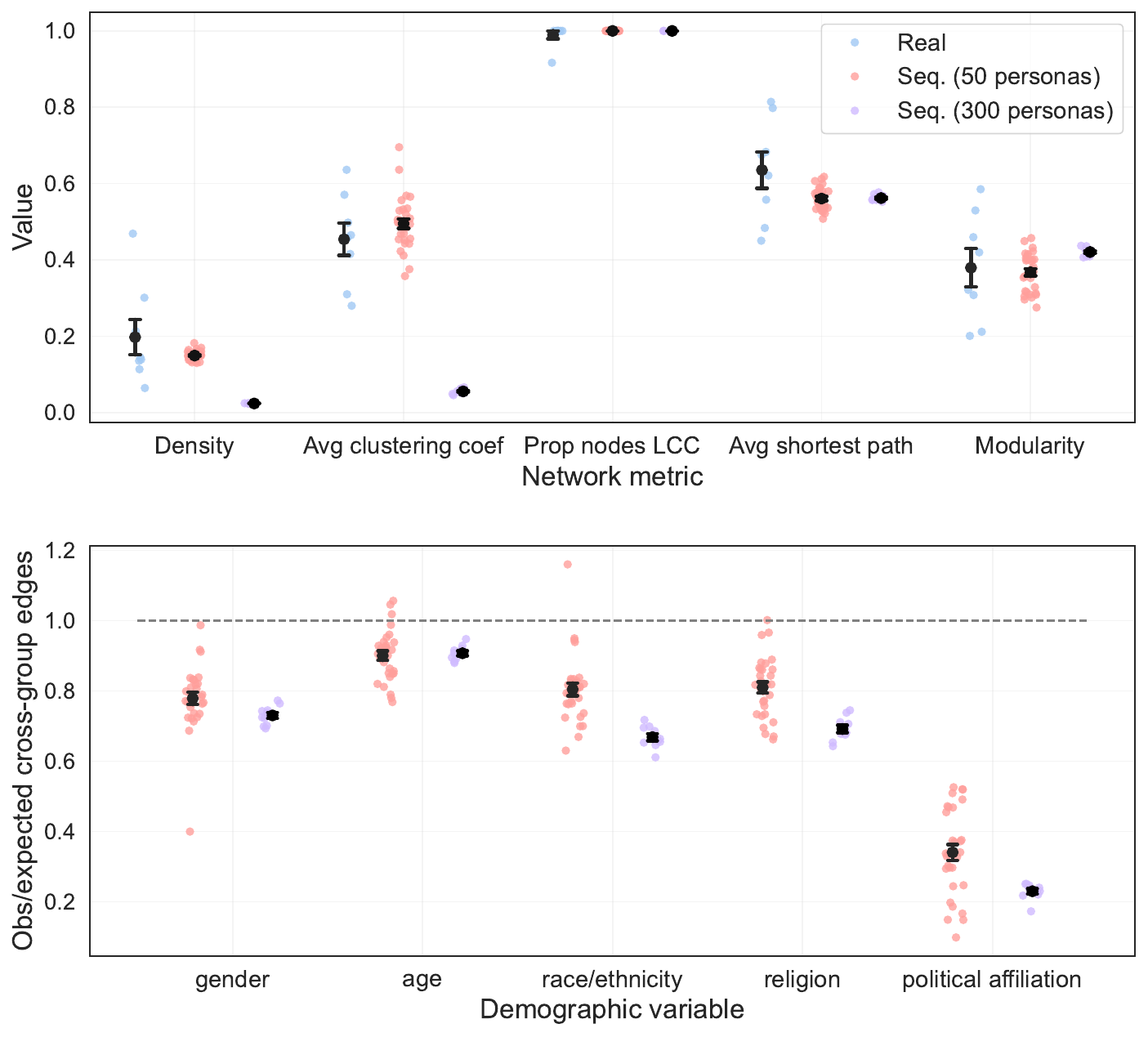}
    \caption{
    Comparing original generated networks with 50 personas to generated networks with 300 personas, where 30 personas are sampled uniformly at random per query.
    \textbf{Top:} structural network metrics. \textbf{Bottom:} homophily, where ratios \rr{below} 1 (marked by the grey line) indicate homophily and \rr{lower} ratios indicate more homophily.
    In both plots, we visualize mean and standard error (in black) and individual data points corresponding to each network. 
    All model results shown here use the Sequential method and GPT-3.5-Turbo.
    }
    \label{fig:networks-300}
\end{figure*}
\rr{To address these tradeoffs, we propose a simple extension of the Local and Sequential methods to make them more scalable.
Previously, when we queried a given persona, we presented it with information about the $N-1$ other personas.
Instead, we could provide it a \textit{subset} of the other personas, with a fixed subset size $k$.
As a result, instead of scaling on the order of $O(N^2D)$, Local and Sequential would scale on the order of $O(ND)$, which is the same as the Global method.}

\rr{To demonstrate how this could be done, we implement a simple version of this extension, where the subset is sampled uniformly at random.
Then, using the Sequential method, we generate networks with 300 personas, sampling 30 random personas per query. Compared to the smaller 50-node networks we generated before, these larger networks show similar levels of homophily across the five demographic variables, and political homophily remains by far the strongest (Figure \ref{fig:networks-300}, bottom). 
In terms of structural characteristics, the larger networks match the smaller ones (and the real networks) on average shortest paths and modularity, but the larger networks have lower density and clustering (Figure \ref{fig:networks-300}, top).
This is expected, due to the sampling: previously, each persona would select friends from 100\% of all other personas, but now each persona selects friends from only 10\% of other personas. Future work could explore methods of non-uniform sampling that can correct for these lower levels of density and clustering, such as using a recommender system-like model (potentially powered by a graph neural network) to choose the subset of other personas that each persona sees.}

\subsection{Evaluating network structure}
\label{sec:app-structure}
In this section, we provide more details about the real social networks that we compared against and define various measures we used to characterize the networks. 
In Table~\ref{tab:real-networks}, we provide basic statistics about the real networks.
\begin{table}[]
    \centering
    \begin{tabular}{c|c|c}
        \textbf{Network} & \textbf{\# nodes} & \textbf{\# edges} \\
        \hline 
        Galesburg & 31 & 63 \\
        Hi-tech & 36 & 91 \\
        Karate & 34 & 78 \\
        Prison & 67 & 142 \\
        Tailor 1 & 39 & 158 \\
        Tailor 2 & 39 & 223 \\
        Moreno freshmen & 31 & 218 \\
        Moreno high school & 70 & 274
    \end{tabular}
    \caption{Summary statistics of the eight real social networks that we use.}
    \label{tab:real-networks}
\end{table}
\paragraph{Real networks.}
We use the following six networks from the CASOS repository \rr{\cite{casos}}:

\textit{Galesburg} \cite{coleman1957diffusion}. This network describes friendship ties between physicians, where they were asked to name three doctors whom they considered personal friends and to nominate three doctors with whom they would discuss medical matters. 
The goal of this study was to analyze the diffusion of a new drug in terms of when physicians first prescribed it, studied in the context of their social network. 

\textit{Hi-tech} \cite{krackhardt1999ties}. This network describes friendship ties between employees of a small hi-tech firm. In a survey, they answered the question, ``Who do you consider to be a personal friend?'' 
Most friendship nominations were reciprocated, and an edge is kept only if both people nominated each other.

\textit{Karate} \cite{zachary1977karate}. This network describes friendships between members of a karate club at a US university.
Due to a schism where the club split into two, this network has often been used to study community structure.

\textit{Prison} \cite{macrae1960sociometric}. This network describes friendship ties between prison inmates. All were asked, ``What fellows on the tier are you closest friends with?'' Each respondent could choose as many or as few friends as he desired.

\textit{Tailor} \cite{kapferer1972tailor}. This network describes relations between workers at a tailor shop in Zambia (then Northern Rhodesia). 
The dataset includes both ``instrumental'' ties (work-related), which we leave out, and ``sociational'' ties (friendship, socioemotional), which we include.
Networks were recorded twice, seven months apart, so we have two networks from this dataset.

We also include two networks from the KONECT repository \rr{\cite{konect}}:

\textit{Moreno freshmen}. This network describes friendship ratings between university freshmen. The edge weights range from -1 (risk of getting into conflict) to +3 (best friend). We keep all edges with strictly positive weight.

\textit{Moreno high school}. This network describes friendship ratings between high school boys. The edge weights range from -1 (risk of getting into conflict) to +3 (best friend). We keep all edges with strictly positive weight.

\paragraph{Network metrics.}
Most of the network metrics that we compare against are straightforward, such as density or average clustering coefficient, which are defined in the main text (Section~\ref{sec:metrics}).
The one more involved metric is modularity, which assesses the quality of a community partition.
Modularity measures the number of edges within the community, compared to how many edges are expected, and it is defined as 
\begin{align}
    Q = \frac{1}{2E} \sum_{ij} ( A_{ij} - \gamma \frac{N_i N_j}{2E} ) \mathds{1}[c_i = c_j], \label{eqn:modularity}
\end{align}
where $\gamma$ is the resolution parameter (set by default to 1) and $c_i$ indicates node $i$'s community in the partition.
As in the main text, $E$ is the total number of edges in the network; $A_{ij}$, as the adjacency matrix, is 1 if nodes $i$ and $j$ are connected and 0 otherwise; and $N_i$ is $i$'s number of neighbors.

We also need to define how we quantified the distance between the generated networks and real networks in Table~\ref{tab:metrics}.
Let $x_1, \cdots, x_m$ represent the values of a metric (e.g., density) from the real networks  (where $m=8$), and let $y_1, \cdots, y_n$ represent the values of the metric from the generated networks (where $n=30)$. 
First, we compute the difference in their mean, divided by the standard deviation of the real network distribution:
\begin{align}
    D &= \frac{|\frac{1}{m}\sum_{i=1}^m x_i - \frac{1}{n}\sum_{j=1}^n y_j|}{\sigma_\textrm{real}}. \label{eqn:norm-mean-diff}
\end{align}
We normalize by standard deviation to make differences comparable across metrics.
We only normalize by the real networks' standard deviation since normalizing by the generated networks' standard deviation would arbitrarily reward higher variance methods.
Second, we use the two-sample Kolmogorov–Smirnov (KS) statistic, which measures the distance between two empirical distributions by comparing their cumulative distribution functions \cite{hodges1958ks}: 
\begin{align}
    F_{\textrm{real}}(u) &= \frac{1}{m} \sum_{i=1}^m \mathds{1}[x_i \leq u] \nonumber \\
    F_{\textrm{gen}}(u) &= \frac{1}{n} \sum_{j=1}^n \mathds{1}[y_j \leq u] \nonumber \\
    D_{KS} &= \sup_u |F_{\textrm{real}}(u)-F_{\textrm{gen}}(u)|. \label{eqn:ks}
\end{align}

\paragraph{Comparing to classical models.}
We compare to several classical models for social network generation, which we describe below, along with how we chose their parameters.
For all models, we set the number of nodes, $N$, to 50, to mimic our LLM experiments with 50 personas.
We use the \texttt{networkx} implementation for all three models.\footnote{\url{https://networkx.org/documentation/stable/reference/generators.html}}

\textit{Erdős–Rényi random graph} \cite{erdds1959random}.
We use the $G_{N,p}$ random graph model, which has $N$ nodes and each edge is included with independent probability $p$.
In our experiments, we simply set $p$ to the average density of the eight real social networks.

\textit{Barabási-Albert preferential attachment} \cite{barabasi1999emergence}.
In this model, one node is added to the graph at each step, and it forms $m$ edges with existing nodes, where each neighbor is sampled with probability proportional to its current degree. 
The \texttt{networkx} implementation starts by default with a star graph on $m+1$ nodes, then adds the remaining $N-m-1$ nodes one at a time. 
Thus, the number of edges in the graph is always $m+((N-m-1) \cdot m)$.
We choose $m=5$, which minimizes the difference between the generated graph's density and the average density of the real social networks.

\textit{Watts-Strogatz small world} \cite{watts1998collective}.
In this model, first a ring is created over $N$ nodes, then each node is joined to its $k$ nearest neighbors, forming a lattice.
Then, with independent probability $p$, each edge $(i, j)$ is rewired, meaning it is replaced with $(i, j')$, where $j'$ is selected from all nodes (aside from $i$ and $i$'s existing neighbors) uniformly at random.
Since each node is joined to its $k$ nearest neighbors, the number of edges in this graph is always $\frac{nk}{2}$.
So, we choose $k=10$, which also minimizes the difference between the generated and real networks' average density.
Then, with $N=50$ and $k=10$, we sweep over possible values of $p$ in $\{0.01, 0.02, \cdots, 0.5\}$ to minimize the difference between the generated and real networks' average clustering coefficient, resulting in $p = 0.15$.

\subsection{Evaluating homophily}
\rr{To evaluate the LLM's level of homophily, we typically use the cross-group ratio in this work (Eq. \ref{eqn:cross-group-ratio}), which measures the ratio of observed-to-expected proportion of cross-group edges.
We also use a closely related measure, the same-group ratio, which measures the ratio of observed-to-expected proportion of same-group edges:
\begin{align}
    H &= \frac{S_{\textrm{obs}}}{S_{\textrm{exp}}} = \frac{\frac{\sum_{ij} A_{ij} \cdot \mathds{1}[g_i = g_j]}{E}}{\frac{\sum_g N_g (N_g - 1)}{N (N-1)}}. \label{eqn:same-group-ratio} 
\end{align}}

To compare the LLM's level of political homophily to prior work, we needed to use their measures of homophily.
First, we define the isolation index, used in \citet{halberstam2016homophily} and \citet{gentzkow2011segregation}.
We define it following \citet{halberstam2016homophily} (Appendix, p. 5).
First, for voter $j \in J$ (they refer to all nodes in their network as voters), let $v_{jC}$ and $v_{jL}$ indicate their number of conservative and liberal followers, respectively.
Then, isolation is defined as
\begin{align}
  \textrm{share-C}_j &= \frac{v_{jC}}{v_{jC} + v_{jL}} \nonumber \\
  \textrm{C-exposure}_i &= \frac{1}{\sum_{j \in J} A_{ij}} \sum_{j \in J} A_{ij} \cdot \textrm{share-C}_j  \nonumber \\
  \textrm{C-exposure}_t &= \frac{1}{N_t} \sum_{i \in I_t} \textrm{C-exposure}_i \nonumber \\
  \textrm{isolation} &= \textrm{C-exposure}_C - \textrm{C-exposure}_L \label{eqn:isolation}
\end{align}
Thus, C-exposure$_i$ for a voter $i$ is the average share-C$_j$ over the voters that they follow, C-exposure$_t$ is the average conservative exposure for voters in group $t$, and isolation is the difference in average conservative exposure for conservative versus liberal voters. 

We also use the polarization measure from \citet{garimella2017twitter}. 
First, they compute the user's leaning, $l$, which is
\begin{align}
  l &= \frac{\alpha}{\alpha + \beta}, \nonumber
\end{align}
where $\alpha$ and $\beta$ indicate how many left-leaning and right-leaning users, respectively, are followed by this user.
Specifically, they begin with a uniform prior, $\alpha = \beta = 1$, then every follow/retweet of a user on each side adds one to that side's parameter ($\alpha$ for left, $\beta$ for right).
Note that their definition of leaning is very similar to the definition of share-C$_j$ in \citet{halberstam2016homophily}, with the addition of the uniform prior.
Then, their definition of polarization $p$ is
\begin{align}
    p &= 2 \cdot |0.5-l|, \label{eqn:polarization}
\end{align}
which lies between 0 and 1, representing how much the user deviates from a balanced leaning of 0.5.
They report average $p$ over users, which is what we measure on our networks.

\end{document}